\documentclass[preprint,prd,aps,showpacs,showkeys,nofootinbib]{revtex4}
\usepackage{txfonts}
\usepackage{stmaryrd}
\usepackage{amsfonts}
\usepackage{amssymb}
\usepackage{}
\usepackage{graphicx}
\usepackage{flafter}

\begin{document}

\title{A New Approach to The Quantum Mechanics}

\author{Yu-Lei Feng$^a$\textasteriskcentered }

\affiliation{$^a$Institute of Theoretical Physics, School of Physics
\& Optoelectronic Technology, Dalian University of Technology,
Dalian, 116024, P. R. China\\}

\begin{abstract}
In this paper, we try to give a new approach to the quantum
mechanics(QM) on the framework of quantum field theory(QFT).
Firstly, we make a detail study on the (non-relativistic)
Schr\"odinger field theory, obtaining the Schr\"odinger equation as
a field equation, after field quantization, the Heisenberg equations
for the momentum and position operators of the particles excited
from the (Schr\"odinger) field and the Feynman path integral formula
of QM are also obtained. We then give the probability concepts of
quantum mechanics in terms of a statistical ensemble, realizing the
ensemble(or statistical) interpretation. With these, we make a
series of conceptual modifications to the standard quantum
mechanics, especially propose a new assumption about the quantum
measurement theory which can solve the EPR paradox from the view of
the QFT. Besides, a field theoretical description to the double-slit
interference experiment is developed, obtaining the required
particle number distribution. In the end, we extend all the above
concepts to the relativistic case so that the ensemble
interpretation is still proper.

Two extra topics are added, in the first one, an operable experiment
is proposed to distinguish the Copenhagen interpretation from the
ensemble one via very different experimental results. While the
second topic concerns with the extensions of the concept of coherent
state to both the Bosonic and Fermionic field cases, to obtain the
corresponding classical fields. And in the concluding section, we
make some general comparisons between the standard QM and the one
derived from the QFT, from which we claim that the QFT is the
fundamental theory.

\end{abstract}


\maketitle

\indent\indent
\section{Introduction}
Quantum theory is well known as one of the most powerful theory in
the last century.  Although it provides an elegant way to describe
the physics of the micro-world, its explanation is so complicated
and obscure, that the Nobel Prize-winning physicist Richard
P.Feynman said that "I can safely say that nobody understands
quantum mechanics"[1].

From the standard point of view, quantum theory includes two parts,
one is the quantum mechanics(QM) which focuses on the behavior of
quantum particles, for instance, the electrons, the photons; the
other part is the quantum field theory(QFT) which gives the rules
for the fields, such as the electromagnetic field. Usually, these
two parts are considered to be independent from each other, by
treating particles and fields as two kinds of independent physical
objects sharing the same quantization scheme. However, from the
QFT[2], it's easy to find that the particles can be treated as
quantum excitations of the corresponding fields, for example, the
electron as excitation of electron field or Dirac field. Since
particles are quantum excitations of fields, then whether QM could
be obtained from QFT seems to be an interesting question[3]. We will
show below that this is possible, and even it provides a new and
natural interpretation to QM.

The paper are roughly divided into three major parts. In the first
one, we study in details a non-relativistic field, the so called
Schr\"odinger field, which is certainly relevant to the ordinary
non-relativistic QM. All of the three standard formula of QM are
obtained from this field theory, the Schr\"odinger equation as field
equation, the Heisenberg equations for the momentum and position
operators of the particles after the field quantization, and the
Feynman path integral formula[4] of QM for particles. Then, in the
second part, the probability concepts of QM are given in terms of a
statistical ensemble, realizing the ensemble interpretation of
QM[5]. With these, we further make a series of conceptual
modifications to the standard quantum mechanics(SQM, the "Copenhagen
Interpretation" ), especially propose a new assumption about the
quantum measurement theory which can solve the EPR paradox from the
view of the QFT. In the end of this part, a field theoretical
description to the double-slit interference experiment is developed,
obtaining the required particle number distribution. In the last
part, an extension to the relativistic QFT is developed, with a
method of separating the particle field from the anti-particle
field, so that the ensemble interpretation is still proper. There
are also two additive topics. In the first one, an operable
experiment is proposed to distinguish the Copenhagen interpretation
from the ensemble one via very different experimental results. While
the second topic concerns with the extensions of the concept of
coherent state for the oscillator to both the Bosonic and Fermionic
field cases, to obtain the corresponding classical fields.

In the concluding section, we make some general analysis on the
basic rules of the standard QM, especially we show that the single
particle operators are not fundamental, but only as derivations of
the QFT. Therefore, we conclude that QFT is the fundamental theory.

\section{Non-relativistic Schr\"odinger field}

The concept of Schr\"odinger field[6] is useful or practical in (low
energy) many-particle physics in which the particle number $N$ is
invariant. Theoretically, this concept is related to the so called
secondary quantization by treating the QM for particles as a
fundamental theory. However, if the field was treated as a basic
element, and the QFT as the fundamental theory, then what would
happen? In this section, we will show the answer to this question.

The action for the Schr\"odinger field can be
\begin{eqnarray}
S=\int dt d^3x
[i\psi^*\dot{\psi}(t,x)-\frac{1}{2m}\nabla\psi^*\nabla\psi(t,x)-\psi^*\psi(t,x)V(x)],
\end{eqnarray}
with $V(x)$ an external potential, for example the Coulomb
potential. For simplicity, we don't consider the field
self-interactions $V(\psi)$ which is necessary in most real physical
situations. In fact, in QFT, relativistic or not, interactions are
already well developed.

From eq.(1), it's easy to find that the field equation is just the
standard Schr\"odinger equation
\begin{eqnarray}
i\frac{\partial\psi}{\partial t}=-\frac{\nabla^2\psi}{2m}+V(x)\psi,
\end{eqnarray}
which can also be obtained from the relativistic equations like the
Dirac equation in the non-relativistic limit. In the SQM, the
Schr\"odinger equation is known as the quantized equation of a
particle, with the wave property. However, in QFT, after the field
quantization, the particles manifest themselves, and satisfy the
field equation automatically.

With the canonical quantization, the Schr\"odinger field will
satisfy the communicative relations
\begin{eqnarray}
[\psi(x),\psi^\dag (y)]_\mp =\delta^3(x-y),
\end{eqnarray}
with the minus for Bosonic case, plus for Fermionic case. In field
theory, we need to consider the space-time symmetries of the
Lagrangian, for example, the symmetry under space-time translation
\begin{eqnarray}
x^\mu\rightarrow x^\mu+a^\mu,
\end{eqnarray}
from which we obtain the energy and momentum operators
\begin{eqnarray}
H=\int d^3x [\frac{1}{2m}\nabla\psi^\dag(x)
\nabla\psi(x)+\psi^\dag(x) V(x)\psi(x)]
\\ P=\int d^3x \psi^\dag(x) (-i\nabla)\psi(x).
\end{eqnarray}
In addition, we can define another two operators
\begin{eqnarray}
X=\int d^3x \psi^\dag(x) x\psi(x)
\\ N=\int d^3x \psi^\dag(x) \psi(x),
\end{eqnarray}
i.e. the position and particle number operators.

Among these operators, there are the following communicative
relations
\begin{eqnarray}
[H,N]=0 ,
\end{eqnarray}
\begin{eqnarray}
[H,P]=i\int d^3x \psi^\dag(x)\nabla V(x)\psi(x) ,
\end{eqnarray}
\begin{eqnarray}
[H,X]=-\frac{1}{m}\int d^3x \psi^\dag(x)\nabla\psi(x) ,
\end{eqnarray}
by using the communicative relations in eq.(3) for both the Bosonic
and Fermionic cases. As is known, $H$ generates the time translation
for an arbitrary operator $O$ constructed with the fields
\footnote{The general form of eq.(12) is actually
$O(t)=e^{iHt}O(0)e^{-iHt}$ in the QFT. }
\begin{eqnarray}
[H,O]=-i\dot O,
\end{eqnarray}
so are those in eqs.(9)-(11). Eq.(9) says that the particle number
is invariant, while the other two are the familiar Heisenberg
equations of the momentum and position operators. Since all these
operators can be represented with the creators and annihilators of
particles in the Fock space, we can denote a single-particle state
as $|1>$, and let eqs.(10) and (11) operate on it, we then have the
QM for single particle. In fact, eqs.(10) and (11) tell us that all
the particles satisfy the QM.

Up to now, we have obtained two main QM equations, one is the
Schr\"odinger equation (2) as field equation, the other is the
system of the Heisenberg equations (10) and (11). Further, we could
obtain the Heisenberg uncertainty relation which is believed to be
the most important property of QM from the communicative relation
\begin{eqnarray}
[X_i,P_j]=i\delta_{ij}\int d^3x \psi^\dag(x)\psi(x) .
\end{eqnarray}
Similarly, we can let it operate on $|1>$ to get the relation for
single particle in QM.

Now we make some studies in the free field case for simplicity
\begin{eqnarray}
\psi(x)=\int \frac{d^3p}{(2\pi)^3} a(p)e^{ipx},  \psi^\dag (x)=\int
\frac{d^3p}{(2\pi)^3} a^\dag(p)e^{-ipx},
\end{eqnarray}
then the (free field) energy and momentum operators become
\begin{eqnarray}
H=\int \frac{d^3p}{(2\pi)^3} E_pa^\dag(p)a(p)=\int
\frac{d^3p}{(2\pi)^3} \frac{p^2}{2m}a^\dag(p)a(p)
\\ P=\int \frac{d^3p}{(2\pi)^3} pa^\dag(p)a(p),
\end{eqnarray}
while the position and particle number operators will be
\begin{eqnarray}
X=\int \frac{d^3p}{(2\pi)^3} a^\dag(p)i\partial_p a(p)
\\ N=\int \frac{d^3p}{(2\pi)^3} a^\dag(p)a(p).
\end{eqnarray}
From eq.(12), we can also have a velocity operator
\begin{eqnarray}
V\equiv \dot{X}=\int \frac{d^3p}{(2\pi)^3} \partial_pE_p
a^\dag(p)a(p)=\int \frac{d^3p}{(2\pi)^3}\frac{p}{m}a^\dag(p)a(p),
\end{eqnarray}
which is similar to the velocity of the non-relativistic particle.
With these operators(or physical observables) obtained, what we then
need are their eigenstates. Obviously, the momentum state can be
easily defined as $|p>=a^\dag(p)|0>$ with normalization
$<p|q>=\delta^3(p-q)$, then what is the position state $|x>$? Let's
define it as follows
\begin{eqnarray}
|x>=\psi^\dag(x)|0>,
\end{eqnarray}
and with eq.(3), it's easy to verify that $|x>$ is just the
eigenstate of operator $X$ with eigenvalue $x$, with normalization
condition
\begin{eqnarray}
<y|x>=<0|\psi(y)\psi^\dag(x)|0>=\delta^3(x-y).
\end{eqnarray}

Now, the Feynman path integral formula of QM can be obtained as
usual in the textbook. With the Heisenberg picture operator
\begin{eqnarray}
X^{H}(t)=e^{iHt}X^{S}(0)e^{-iHt},
\end{eqnarray}
and a moving basis $|x,t>$ which satisfies
\begin{eqnarray}
X^{H}(t)|x,t>=x|x,t>,
\end{eqnarray}
we then have
\begin{eqnarray}
|x,t>=e^{iHt}|x>=e^{iHt}\psi^{\dag
S}(x)e^{-iHt}e^{iHt}|0>=\psi^{\dag H}(x,t)|0>.
\end{eqnarray}
Thus the transition amplitude is
\begin{eqnarray}
<x_2,t_2|x_1,t_1>=<x_2|e^{-iH(t_2-t_1)}|x_1>=<x_2|e^{-i\hat{H}(t_2-t_1)}|x_1>,
\end{eqnarray}
with $\hat{H}$ the familiar single-particle Hamiltonian. Obviously,
eq.(25) is the starting point of Feynman path integral formula of
QM.

In fact, the Feynman path integral formula can be derived in a pure
field theoretical way as follows. From eq.(24), eq.(25) could be
rewritten as
\begin{eqnarray}
<x_2,t_2|x_1,t_1>=<0|\psi^{H}(x_2,t_2)\psi^{\dag H}(x_1,t_1)|0>,
\end{eqnarray}
which is just the propagator of the field in QFT[2]. Now we can
compute this propagator with the standard field method, i.e. define
the interaction picture
\begin{eqnarray}
\psi^{\dag I}(x,t)=e^{iH_0t}\psi^{\dag S}(x)e^{-iH_0t},
\end{eqnarray}
with the free field Hamiltonian $H_0$ constructed from the fields in
the Schr\"odinger picture. Then we have the following relation
\begin{eqnarray}
\psi^{\dag H}(x,t)=e^{iHt}e^{-iH_0t}\psi^{\dag
I}(x,t)e^{iH_0t}e^{-iHt}=U^{\dag}(t,0)\psi^{\dag I}(x,t)U(t,0),
\end{eqnarray}
where the time-evolution operator $U(t,0)=e^{iH_0t}e^{-iHt}$
satisfying the equation
\begin{eqnarray}
i\frac{\partial U(t,0)}{\partial t}=H_{int}^{I}(t)U(t,0),
\end{eqnarray}
with a time-ordering formal exponential solution
\begin{eqnarray}
U(t,0)=T\exp[-i\int_{0}^{t} dt'H_{int}^{I}(t')].
\end{eqnarray}
With eq.(28), the right-hand-side of eq.(26) will become
\begin{eqnarray}
<0|U^{\dag}(t_2,t_0)\psi^{
I}(x_2,t_2)U(t_2,t_0)U^{\dag}(t_1,t_0)\psi^{\dag
I}(x_1,t_1)U(t_1,t_0)|0>,
\end{eqnarray}
for a general reference time $t_0$. With the condition
$U(t_1,t_0)|0>=|0>$ for the Schr\"odinger field, and the following
property of $U$[2]
\begin{eqnarray}
U(t_2,t_0)U^{\dag}(t_1,t_0)=U(t_2,t_1),
\end{eqnarray}
eq.(31) will be further simplified as
\begin{eqnarray}
<0|\psi^{I}(x_2,t_2)U(t_2,t_1)\psi^{\dag
I}(x_1,t_1)|0>=<0|T\psi^{I}(x_2,t_2)\psi^{\dag
I}(x_1,t_1)e^{-i\int_{t_1}^{t_2} dt'H_{int}^{I}(t')}|0>.
\end{eqnarray}
To compute eq.(33), the element is the propagator
$K(x_2,t_2;x_1,t_1)$ of free field defined as
\begin{eqnarray}
K(x_2,t_2;x_1,t_1)=<0|\psi^{I}(x_2,t_2)\psi^{\dag I}(x_1,t_1)|0>.
\end{eqnarray}
With eqs.(14) and (27), the free propagator is
\begin{eqnarray}
K(x_2,t_2;x_1,t_1)=\int \frac{d^3p}{(2\pi)^3}
e^{-iE_p(t_2-t_1)+ip(x_2-x_1)},
\end{eqnarray}
which can be solved by the Gaussian integral formula, and we thus
have
\begin{eqnarray}
K(x_2,t_2;x_1,t_1)=[\frac{m}{2\pi i(t_2-t_1)}]^{\frac{3}{2}} \exp[i
\frac{m}{2}(\frac{x_2-x_1}{t_2-t_1})^2(t_2-t_1)].
\end{eqnarray}
Furthermore, we can infer from eq.(34) that
\begin{eqnarray}
K(x_3,t_3;x_1,t_1)=\int d^3x_2K(x_3,t_3;x_2,t_2)K(x_2,t_2;x_1,t_1),
\end{eqnarray}
where the completeness relation
\begin{eqnarray}
\int d^3x\psi^{\dag I}(x,t)|0><0|\psi^{I}(x,t)=\int d^3x|x><x|=I,
\end{eqnarray}
of the Schr\"odinger field has been used. Now just like the case in
the standard derivation of the path integral formula, the time
interval $(t_2,t_1)$ can be split up to many small slices
$\epsilon$, for example $N$, then eq.(36) can be rewritten as
\begin{eqnarray}
K(x_2,t_2;x_1,t_1)=[\frac{m}{2\pi i\epsilon}]^{\frac{3N}{2}}\int
\prod_id^3x_i\exp[i
\sum_i\frac{m}{2}(\frac{x_{i+1}-x_i}{\epsilon})^2\epsilon].
\end{eqnarray}
Eq.(33) can be computed in perturbative power series, in which the
zero order is just the free filed propagator in eq.(34), while the
first order is
\begin{eqnarray}
(-i)\int_{t_1}^{t_2} dtd^3x K(x_2,t_2;x,t)V(x)K(x,t;x_1,t_1),
\end{eqnarray}
similar for larger order. In order to compare with the path integral
formula, we give here the standard path integral formula[2]
\begin{eqnarray}
\lim_{N\to \infty}[\frac{m}{2\pi i\epsilon}]^{\frac{3N}{2}}\int
\prod_id^3x_i\exp i
\sum_i[\frac{m}{2}(\frac{x_{i+1}-x_i}{\epsilon})^2\epsilon-\epsilon
V(\frac{x_{i+1}+x_i}{2})].
\end{eqnarray}
Obviously, we only need to compare the potential terms, which can
also be rewritten order by order. The zero order is the free case,
just like eq.(39), while the first order is
\begin{eqnarray}
(-i)[\frac{m}{2\pi i\epsilon}]^{\frac{3N}{2}}\int \prod_id^3x_i\exp
i[\sum_i\frac{m}{2}(\frac{x_{i+1}-x_i}{\epsilon})^2\epsilon]\sum_i\epsilon
V(\frac{x_{i+1}+x_i}{2}),
\end{eqnarray}
which in the large $N$ limit is just eq.(40) with the sum over the
positions and time slices replaced with the integral
$\int_{t_1}^{t_2} dtd^3x$. Now, let's see the second order, for the
field case, we have
\begin{eqnarray}
\frac{(-i)^2}{2}\int_{t_1}^{t_2} dtd^3x\int_{t_1}^{t_2} dt'd^3x'
K(x_2,t_2;x,t)V(x)K(x,t;x',t')V(x')K(x',t';x_1,t_1),
\end{eqnarray}
while from eq.(41), we have
\begin{eqnarray}
(-i)^2\sum_i\epsilon^2
V^2(\frac{x_{i+1}+x_i}{2})+\frac{(-i)^2}{2}\sum_{i\neq j}\epsilon^2
V(\frac{x_{i+1}+x_i}{2})V(\frac{x_{j+1}+x_j}{2}),
\end{eqnarray}
where the second term is easily to be identified, then what about
the first one? Noting that from eq.(34), we have
\begin{eqnarray}
K(x_2,t;x_1,t)=<0|\psi^{I}(x_2,t)\psi^{\dag
I}(x_1,t)|0>=\delta^3(x_2-x_1),
\end{eqnarray}
this means that the first term in eq.(44) can be easily from the
$t\to t'$ limit of eq.(43). Then order by order, we can find that
eqs.(33) and (41) are actually identical, confirming the eq.(25)
which give the path integral formula of QM from the Schr\"odinger
field theory.

Up to now, we have obtained all the three equivalent approaches to
the non-relativistic QM on the framework of the quantum
Schr\"odinger field theory, with all the physical observables
completely constructed with the quantized fields, especially the
position operator in eq.(7) whose meaning is obscure for field. In
the next section, we will show some further physical results of the
Schr\"odinger field theory, especially the possible modifications to
the SQM.

\section{MODIFICATIONS TO THE SQM }
\subsection{Probability Concepts From QFT}

Since the particle number is invariant as in eq.(9), the statistical
property of some collection of particles in the field theory could
be transferred to the probability property of single particle.

Supposing that $\psi(t,x)$ has the energy expansion\footnote{For
simplicity, we assume that there is no state degeneration and the
expansion eq.(46) can be treated as either the wave function or the
field.}
\begin{eqnarray}
\psi(t,x)=\sum_{n}a_n\psi_n(x)\exp(-iE_n t),
\end{eqnarray}
from it, we can obtain that, in QM, the probability for a particle
to be at state $n$ is\footnote{Here, the time is ignored because
they are stationary states, the same below.}
\begin{eqnarray}
P_n\equiv \frac{|\int d^3x\psi^*_n(x)\psi(x)|^2}{\int
d^3x\psi^*(x)\psi(x)}=\frac{|a_n|^2}{\int d^3x\psi^*(x)\psi(x)}.
\end{eqnarray}
A possible field theoretical generalization could be \footnote{For
the continuous case like the energy density, we have the probability
density $P(\omega)d\omega\equiv
\frac{<a^\dag(\omega)a(\omega)d\omega>}{<\int d\omega
a^\dag(\omega)a(\omega)>}$.}
\begin{eqnarray}
P_n\equiv \frac{<\int d^3x\psi^\dag(x)\psi_n(x)\int
d^3y\psi^*_n(y)\psi(y)>}{<\int d^3x\psi^\dag(x)\psi(x)>}=
\frac{<a^\dag_{n}a_n>}{<\sum_na^\dag_{n}a_n>},
\end{eqnarray}
with notation $>$ standing for particle state in Fock
space\footnote{In fact, as we will show below, the particle state
$>$ is usually standing for some ensemble, i.e. copies of a single
particle or many-particle systems conceptually.}. It seems that
eqs.(47) and (48) could be identical, with the meaning that, among a
collection of particles excited from the field, \emph{the
probability for the particle picked up to be at state $n$ is in fact
the ratio of the particle number at this state by the total particle
number in this collection.} With this identification, the mean value
of some physical quantity, for example, the energy, is
\begin{eqnarray}
\bar{E}\equiv \frac{<H>}{<\int d^3x\psi^\dag(x)\psi(x)>}=
\frac{<E_na^\dag_{n}a_n>}{<\sum_na^\dag_{n}a_n>},
\end{eqnarray}
with $H$ the field energy operator defined in eq.(5).

Obviously, eqs. (48)and (49) are more familiar to us conceptually,
based on the traditional probability concepts which come from the
statistical property of a collection of particles for single
particle or collections of a collection of particles for
many-particle system, i.e. some statistical ensemble of systems,
made up of the particles excited from the field. However, in order
for the identity for eqs.(47) and (48), we have to make an important
assumption that, \emph{the state of each single particle is definite
and unique (but unknown to us if without any measurement), as long
as there's no disturb}, which is very different from that in SQM.
This is an extension of the Newton's first law with the velocity
replaced by the state. Then the state $>$ is usually standing for
some ensemble made up of particles with different kinds of states so
that eq.(48) is proper.

One important example of the above ensemble is a sample of particles
in an experiment about some physical process, in other words, all
the particles in the sample come from a physical process such as a
scattering. It is certain that the sample must contain some
information about the physical process, for example the scattering
angle distribution, which can be studied by the ratios of the
particles in all the angles. And what the scattering distribution
can tell us is just the probability for one particle to be observed
in one angle, which can be solved by QM or QFT. Therefore, in this
sense, the QM and QFT should be physically identical. Since the QM
could be obtained from QFT as shown in the last section, the
probability concept could be realized with the use of the sample or
ensemble specific for some property of the particle, for instance,
the scattering angle distribution.

Now, let's consider a simple example. Suppose that in the SQM, there
is a state vector\footnote{Here $|1>=a_1^\dag|0>$, and we have
assigned an arbitrary phase term which could be space-time
dependent, and we will see below that this important phase term
usually comes from interactions, such as a measurement.}
\begin{eqnarray}
|\phi>=\sqrt{\frac{1}{3}}|1>+e^{i\delta(t,x)}\sqrt{\frac{2}{3}}|2>.
\end{eqnarray}
If this was a state for a single particle, it would be a so called
pure state in SQM, and the density matrix is
\begin{eqnarray}
|\phi><\phi|=\frac{1}{3}|1><1|+\frac{2}{3}|2><2|+\frac{\sqrt{2}}{3}(e^{-i\delta(t,x)}|1><2|+e^{i\delta(t,x)}|2><1|).
\end{eqnarray}
The wave function for this state can be resulting
from\footnote{Noting that by substituting $|\phi>$ into eq.(48), we
will obtain the form of eq.(47), and state in eq.(53) below has the
same physical results as $|\phi>$.}
\begin{eqnarray}
<0|\psi(x)|\phi>=\sqrt{\frac{1}{3}}\psi_1(x)+e^{i\delta(t,x)}\sqrt{\frac{2}{3}}\psi_2(x),
\end{eqnarray}
with the use of field expansion in eq.(46). Then from eq.(47), this
state says that, the probability for this particle to be at state 1
is $\frac{1}{3}$, and to be at state 2 is $\frac{2}{3}$. According
to the above discussions, the probability properties of single
particle could be from the statistical properties of a collection of
particles. Then, there must be a collection of particles which
consists of copies of a single particle
\begin{eqnarray}
[\underbrace{ 1,1,\cdots 1 }_{\frac{N}{3}};\underbrace{ 2,2,\cdots 2
}_{\frac{2N}{3}}].
\end{eqnarray}
The meaning is obvious, i.e. within the $N$ particles, $\frac{N}{3}$
are at state 1, the rest are at state 2, i.e. an ensemble for the
single particle\footnote{Do not confuse with the many-particle state
which is a real N-particle system. In fact, it's easy to distinguish
them by noting that it's impossible to include so many particles
with the same state for the Fermionic case.}. And the density matrix
for single particle should be
\begin{eqnarray}
\rho=\frac{1}{3}|1><1|+\frac{2}{3}|2><2|.
\end{eqnarray}
Comparing with eq.(51), the off-diagonal terms disappear, which is
one of the important distinctions between our statistical ensemble
and the SQM. In fact, this example involves the problem of
superposition state which will be discussed in details in the next
subsection, here what we only need to know is that the state in
eq.(50) would hardly be a pure state for a single particle, but
should be identified with a so called mixed state in SQM, with
density matrix in eq.(54).

In fact, if we "normalize" the field function, a possible field
expression may be as follows
\begin{eqnarray}
\phi^\dag(x)|0>=\sqrt{\frac{1}{3}}\psi^*_1(x)|1>+\sqrt{\frac{2}{3}}e^{-i\delta(t,x)}\psi^*_2(x)|2>,
\end{eqnarray}
with the "re-normalized" field
\begin{eqnarray}
\phi^\dag(x)=\sqrt{\frac{1}{3}}a_1^\dag\psi^*_1(x)+\sqrt{\frac{2}{3}}e^{-i\delta(t,x)}a_2^\dag\psi^*_2(x),
\end{eqnarray}
which incorporates both the field and wave function properties. Then
let's consider the following expression
\begin{eqnarray}
\int d^3x\phi^\dag(x)|0><0|\phi(x),
\end{eqnarray}
which is formally the intermediate section of two propagators
\begin{eqnarray}
\int dy_0
d^3y<0|\psi(x_0,x)\psi^\dag(y_0,y)|0><0|\psi(y_0,y)\psi^\dag(z_0,z)|0>.
\end{eqnarray}
With the orthonormalization conditions, eq.(57) is just the density
matrix in eq.(54)\footnote{The density matrix formula in eq.(57) has
a time evolution $\rho(t)=e^{iHt}\rho(0)e^{-iHt}$, different from
the one in SQM, because it's made up of fields. }! All these confirm
our ideas above. We can also obtain the mean value of a physical
quantity, for example, the energy
\begin{eqnarray}
\bar{E}=Tr(\rho H)=\sum_n<n|\int
d^3x\phi^\dag(x)|0><0|\phi(x)H|n>=\frac{1}{3}E_1+\frac{2}{3}E_2.
\end{eqnarray}
Considering an interaction $B(x)$, we can also have a general
density matrix
\begin{eqnarray}
\int d^3x\psi^\dag(x)|0>B(x)<0|\psi(x)=\sum_{mn}B_{mn}|m><n|,
\end{eqnarray}
with $B_{mn}$ the transition amplitude
\begin{eqnarray}
B_{mn}=\int d^3x\psi_m^*(x)B(x)\psi_n(x).
\end{eqnarray}

Here is a note about $N$, the particle number in the single particle
ensemble. From the above example, it appears that $N$ could be any
number, and it's indeed so. The reason is that the ratios of
particles with different kinds of states are almost fixed for an
ensemble corresponding with some physical process, and just like the
law of large numbers in probability theory, we need to take $N\to
\infty $ in the real situations. The above ideal example is only for
showing how to transform the ensemble (53) into the QM-like
formalism in eq.(55)or (56), i.e. "re-normalize" the field
functions, and obtain the density matrix for a single particle in
eq.(57).

The above single-particle ensemble is a simple one with some fixed
probabilities. A general one can be described as follows, treating
the particle as one system, and we don't know the exact state of
this system without any measurement, so we have to list all the
possibilities, i.e. the particles at state 1, or 2, or \ldots etc.
For every possibility, there will be a corresponding (variable)
probability(i.e. the ratios of the particle numbers), $P_1$, $P_2$,
\ldots etc. Then \emph{a state for the ensemble(or ensemble state)}
will be
\begin{eqnarray}
e^{i\delta_1}\sqrt{P_1}|1>+e^{i\delta_2}\sqrt{P_2}|2>+\cdots+e^{i\delta_n}\sqrt{P_n}|n>+\cdots=\sum_n\alpha_n|n>,
\end{eqnarray}
an extension of eq.(50), and easily to see it has the same form with
a general state vector in SQM. If we would like to know the exact
state of the particle, we have to observe it, and obtain that the
particle is in fact at some definite state, for example, $k$. It
appears that there is the so called quantum collapse here, as in
SQM, but no!

Noting that the state in eq.(62) is for the single-particle
ensemble, not for the particle, this means that there must be a
realization to the ensemble, and the simple example in eq.(50) or
(53) is one with definite probabilities for each state. We could
also realize the ensemble artificially by collecting particles with
arbitrary unknown ratios at will. And no matter whether the
probabilities are already known(fixed) or unknown, the essential
feature is the same. The collapse due to an observation on the state
in eq.(62) can be explained with a familiar example. Supposing there
are three balls with red or blue colors in a bag, and further we
know that there are one red ball and two blue ones. Then from these,
we know that if we pick a ball arbitrarily, it can be red or blue,
and the probability for it to be red is $\frac{1}{3}$, $\frac{2}{3}$
for blue, then this "state" about the color of a single ball could
be described by the state in eq.(50). However, if we observe the
ball and find that it is blue, then how to explain this observation?
Is there also a collapse classically? The only reason is that the
state in eq.(50) is an ensemble state, which is just a useful tool,
and QM or QFT is a realization of it physically by collecting the
copies of the particle conceptually. The reason for using an
ensemble is that \emph{we can't obtain the exact information of a
system without any measurement and can only list all the
possibilities with the corresponding probabilities}. Therefore, to a
certain extent, QM is much more consistent with the ensemble
interpretation.

We can extend the single-particle ensemble to a general N-particle
one which is usually seen in the statistical mechanics or
many-particle physics. Considering $N$(identical) particles within
$k$ states (without degenerations), we can start with
\begin{eqnarray}
\psi^\dag(x_1)\psi^\dag(x_2)\cdots \psi^\dag(x_N)|0>,
\end{eqnarray}
up to some constant, and it can also be rewritten in Fock space with
a recombination as, for example the Bosonic case
\begin{eqnarray}
\sum
_{n_1+n_2+\cdots+n_k=N}\sqrt{\frac{P_{[n_j]}}{S}}e^{i\delta_{[n_j]}}(a_1^\dag)^{n_1}(a_2^\dag)^{n_2}\cdots
(a_k^\dag)^{n_k}|0>,
\end{eqnarray}
with $S$ a symmetry factor which is $(n_1)!(n_2)!\cdots(n_k)!$ for
Bosonic case and $1$ for the Fermionic case, $P_{[n_j]}$ the
probability for a possible distributions $[n_j]$ with $n_1$
particles in state 1, $n_2$ particles in state 2, etc, specially
$n_j=0,1$ for the Fermionic case due to the Pauli exclusion
principle, and the sum is over all the possibilities. With some
other extra specific conditions for bosons and fermions, we could
further obtain the Bose and Fermi statistics. It's easy to see that
eq.(64) is just a state for an N-particle ensemble, the extension of
eq.(62). And the reason for using an ensemble is the same with the
single-particle case, i.e. we don't know the states for all the
particles to specify the state of the system without measurements.
Further, in this case, the situation is much more complicated than
the single-particle case because of the large number of the
particles and interactions among them. Therefore, we can only
describe the system with the method of statistical mechanics by
finding out the most possible distribution in eq.(64).

In order to understand these, let's take a look at the above simple
example in eq.(53) again, with $N=3$, that is three particles within
two states. We  still take the Bosonic case, and it's easy to see
that there are four possibilities with density matrixes
\begin{eqnarray}
\rho_1=|1><1| \\ \rho_2=\frac{1}{3}|1><1|+\frac{2}{3}|2><2| \\
\rho_3=\frac{2}{3}|1><1|+\frac{1}{3}|2><2| \\ \rho_4=|2><2| .
\end{eqnarray}
Recall that that example is originally a single particle ensemble,
so the above four density matrixes are all for single particle, that
is we can only obtain the information about single particle from
them. While the density matrixes for a real 3-particle system, from
which we could obtain the information about the whole system, are
respectively(up to some normalization constants)
\begin{eqnarray}
\sigma_1=(a_1^\dag)^3|0><0|(a_1)^3 \\ \sigma_2=a_1^\dag(a_2^\dag)^2|0><0|(a_2)^2a_1 \\
\sigma_3=a_2^\dag(a_1^\dag)^2|0><0|(a_1)^2a_2 \\
\sigma_4=(a_2^\dag)^3|0><0|(a_2)^3 ,
\end{eqnarray}
corresponding to the ensemble state in eq.(64). If the four
possibilities in eqs.(65)-(68) have equal probability\footnote{This
is only an assumption, and for large $N$ we could obtain the most
possible distribution from the statistical mechanics. }, i.e. $1/4$,
then the final result for single particle is
\begin{eqnarray}
\rho=\frac{1}{4}(\rho_1+\rho_2+\rho_3+\rho_4)=\frac{1}{2}(|1><1|
+|2><2|),
\end{eqnarray}
which can also obtain via the ordinary probability computations. In
fact, there exists a class of N-particle ensemble which can be made
up with the single-particle ones for each particle, which can be
seen from eq.(63), with each field function $\psi^\dag(x_j)$
standing for a single-particle ensemble, i.e. the "re-normalized"
field in the form of eq.(56). Then the probability $P_{[n_j]}$ is
the multiplication of the corresponding probabilities of single
particle ensembles and some symmetry factors.

In SQM, the state in eq.(50) is a superposition state, which is
related with the principle of superposition of states. However, from
the view of ensemble, the state in (50) is not physical, but a state
for an ensemble(or mixed state in SQM), so is the one in eq.(62).
Therefore, it appears that the principle of superposition is
suitable for the field, in other words, the collection of particles
with different states i.e. the ensemble. We will make some detail
discussions on these in the next subsection, which give some
modifications to the SQM.

\subsection{Conceptual Modifications to The SQM}

Based on the field theoretical descriptions above, in this
subsection, we will give a series of modifications to the SQM, and
make some detail discussions on the superposition principle and
quantum measurement theory.

(1)\emph{The wave function, or the probability amplitude in QM is
not a fundamental element, but a derivation of the field
$\psi(t,x)$, a distribution in space-time, which is real in nature.
In addition, the original Schr\"odinger equation is the
non-relativistic field equation, as shown in eq.(2).}

This is the result of section II., where the Schr\"odinger equation
(wave form) and Heisenberg equations (particle form) are both
derived from the field theory. The meaning for the latter is clear,
while the Schr\"odinger equation is ambiguous because in SQM it's
rewritten in the following form
\begin{eqnarray}
i\frac{\partial}{\partial t}|\phi(t)>=\hat{H}|\phi(t)>,
\end{eqnarray}
with $\hat{H}$ the Hamiltonian operator for single particle. What
this equation can tell us is the evolution of the state of the
particle, somewhat deterministic, that is given the state at some
time $t_0$, it could determine the state thereafter. However, the
original Schr\"odinger equation (2) can also be considered to be the
time evolution of the quantized field
\begin{eqnarray}
i\frac{\partial}{\partial t}\psi(t,x)=-[H,\psi(t,x)],
\end{eqnarray}
with $H$ the Hamiltonian of the field as in eq.(5), or in a more
compact form
\begin{eqnarray}
\psi(t,x)=e^{iHt}\psi(0,x)e^{-iHt}.
\end{eqnarray}
Thus, with a state vector $|\phi>$, and by using eqs.(20) and (52),
we have the wave function
\begin{eqnarray}
\phi(t,x)=<0|\psi(t,x)|\phi>=<0|e^{iHt}\psi(x)e^{-iHt}|\phi>=<x|\phi(t)>,
\end{eqnarray}
with
\begin{eqnarray}
|\phi(t)>=e^{-iHt}|\phi>\rightarrow e^{-i\hat{H}t}|\phi>,
\end{eqnarray}
where we have reduced the field Hamiltonian $H$ into single particle
one $\hat{H}$ because of the state $<x|$ in eq.(77), just like the
case in eq.(25). Therefore, field equation (75) is much more
fundamental than the state evolution equation (74).

(2)\emph{The probability $\int d^{3}x\phi^*\phi(t,x)$ in SQM
corresponds to the particle number operator $\int
d^{3}x\psi^\dag\psi(t,x)$, then the probability conservation in the
SQM is in fact the conservation of total particle number in
non-relativistic QFT as shown in eq.(9).}

As described in the last subsection, the probability concepts for
single particle comes from the statistical concepts of a statistical
ensemble like the one in eq.(53) or (62). And in order for the
identity of eqs.(47) and (48), we  have made an important
assumption, \emph{the state of every single particle is definite and
unique (but unknown to us if without any measurement)}\footnote{We
could know the state only if we had observed it, i.e. interacted
with it.}. Of course, this is very different from the assumption in
SQM, where the state for single particle could be the superposition
of states in the following form
\begin{eqnarray}
|\phi>=\sum_{n}\alpha_n|n>,
\end{eqnarray}
which is the same form as the ensemble state in eq.(62). Because
eq.(74) is not a fundamental equation, then the state in eq.(79)
loses its physical meaning as a state of single particle, so does
the wave function $\phi(t,x)$. All these involve the so called "the
superposition principle" in QM, and now let's see how to interpret
it properly.

In SQM, the superposition principle generally says that, if $\psi_1$
and $\psi_2$ are both the states of a system, then the linear
combination $\alpha\psi_1+\beta\psi_2$ (with $\alpha$ and $\beta$
arbitrary complex numbers) is also a possible state of the system.
This principle can be proved loosely by the linearity of the
Schr\"odinger equation. It also could be seen roughly from the
expansion (46) or (79), which could be interpreted to be
superposition for single particle state in SQM. But the field
theoretical form (46) may also be interpreted as followed, the
states are all the possibilities for the particles excited from the
field, and since we have assumed the definite and unique for single
particle, the concept of ensemble is needed. Then we could have the
following modification

(3)\emph{The superposition principle is suitable for the field, that
is an ensemble of particles, or ensemble states in eqs.(62) and
(64), not for a single particle.}

In fact, the superposition principle in QM is so strongly dependent
on the linearity of the Schr\"odinger equation that if we include
the self-interaction terms into the action in eq.(1), the resulting
equation is non-linear and hard to solve, and the expansion in
eq.(46) is useless, and we could only use the free theory expansion
to obtain the perturbative power series for the interactions, as
shown in eq.(33). Therefore, we could not decide which principle
(SQM's or ours) is much more physical, because it's necessary to
combine the assumption of the quantum measurement. As is known to
us, in SQM, there is the so called mysterious \emph{quantum
collapse} owing to the superposition principle. However, there is
nothing abnormal with our ensemble concepts, which will be discussed
in the next subsection\footnote{We have already a simple explanation
below eq.(62) in the last subsection. }.

In the rest of this section, we will introduce some examples about
the superposition principle suitable for both the SQM and QFT
mathematically. First of all, let's distinguish two concepts,
\emph{superposition state} and \emph{superposition of
states}\footnote{These concepts may be different from those in SQM,
but the discussions below will be self-consistent. }. Easily to see,
the former is included in the latter. In fact, superposition of
states can be generally expressed mathematically as
$\alpha\psi_1+\beta\psi_2+\cdots$. However, the superposition state
as state of single particle must be physical in nature, though
mathematically has the form of superposition of states. In other
words, a physical state $\psi$ which can be expressed as
\begin{eqnarray}
\psi=\alpha\psi_1+\beta\psi_2,
\end{eqnarray}
with some $\mathbf{fixed}$ numbers $\alpha$ and $\beta$(up to some
overall normalization constant) to specify the physical properties,
and this mathematical expression is just a convenient relation for
some analysis.

Let's introduce a class of superposition state. The first example is
the eigenstates of momentum $e^{ip\cdot x}$ which can be expressed
mathematically as the combinations of some special functions like
spheric harmonics functions(Rayleigh expansion), and vice versa. In
QM language, these are the transformations between momentum and
angular momentum states
\begin{eqnarray}
\{|p_1,p_2,p_3>\}\leftrightharpoons \{|p,l,m>\}.
\end{eqnarray}
The second example will be used in the next subsection. It involves
the spin states of electrons, i.e. $|\uparrow_z>$ and
$|\downarrow_z>$, and for any direction
$\hat{n}=(\sin\theta\cos\varphi,\sin\theta\sin\varphi,\cos\theta)$,
$|\uparrow_n>$ and $|\downarrow_n>$. There are transformations
between them , for example,
\begin{eqnarray}
|\uparrow_n>=e^{-i\frac{\varphi}{2}}\cos\frac{\theta}{2}|\uparrow_{z}>+e^{i\frac{\varphi}{2}}\sin\frac{\theta}{2}|\downarrow_{z}>.
\end{eqnarray}
with fixed coefficients to specify the direction $\hat{n}$, thus, we
have similarly
\begin{eqnarray}
\{|\uparrow_z>,|\downarrow_z>\}\leftrightharpoons
\{|\uparrow_n>,|\downarrow_n>\}.
\end{eqnarray}
The last example is about the coupling of angular momenta. In order
to understand it, we first give a simple example which involves the
same essential feature with the coupling of angular momenta. Suppose
that there are two particles 1 and 2 with states $\{(E_1,p_1)\}$ and
$\{(E_2,p_2)\}$ respectively. Easily to see, this description is
proper in the lab frame, and we can also describe them in the
center-of-mass frame with states $\{(E_c,P),(E,p)\}$, representing
the energy of the center-of-mass frame, total momentum, relative
energy and momentum, respectively. Then there should be some
relations for these two classes of states as eqs.(81) and (83). In
fact, the coupling of angular momenta operates similarly, and we
will have the following correspondence for two angular momenta
$J_1,J_2$
\begin{eqnarray}
\{|j_1,m_1;j_2,m_2>\}\leftrightharpoons \{|j,m>\},
\end{eqnarray}
with the CG coefficients in the relations between them.

Obviously, the above three examples satisfy the conditions of
superposition state, i.e. physical and superposition of states. The
reason for putting them into one class is that there are states
transformations for all of them, i.e. eqs.(81),(83) and (84).
Furthermore, we can see that all these transformations are related
to some coordinate transformations, (81) for Cartesian coordinate
and spherical coordinate, (83) for rotations on a sphere, and (84)
for transformations between lab frame and center-of-mass
frame\footnote{With frame transformation
$X=\frac{m_1x_1+m_2x_2}{m_1+m_2}$.}. And according to the Wigner
theorem, all these transformations between the states are all
unitary with $\mathbf{fixed}$ coefficients(up to some overall phase
terms), the last key condition for the superposition state. In fact,
this class of superposition state can be defined for any two
complete states, for instance $\{|n>\}\leftrightharpoons \{|i>\}$,
with the relations $|n>=\sum_i|i><i|n>$ using the completeness
relation $I=\sum_i|i><i|$, vice versa. Obviously, these satisfy
those conditions of the superposition state, and with eq.(84), it's
possible for states of any many-particle system, as long as they
have some definite quantum numbers, that is they are physical
states.

Now let's see the state in eq.(50) again, and easily to see, it's
superposition of states. However, if it was also a superposition
state, then what its quantum numbers are? Further, the coefficients
are not fixed, especially the possible arbitrary phase difference
$\exp(i\delta)$ between them. Therefore, this state is not a
physical one for single particle, just an ensemble state. The same
things happen to the state in eq.(62) which describes a general
statistical ensemble. Therefore, we can conclude that the ensemble
state as in eqs.(62) and (64) are not superposition states, only
superposition of states.

In one word, in our familiar examples, superposition state appears
only in some cases like eqs.(81),(83) and (84). And the expressions
like eq.(82) are only the mathematical relations between the
corresponding states. When there are interactions, the coefficients
of those expressions would obtain some arbitrary phase terms, then
the condition of fixed coefficients is broken, and the superposition
state will change into ensemble states. All these will be shown in
the next subsection, where the interaction is the quantum
measurement, and then which superposition principle(SQM's or QFT's)
is much more proper will also be clear.

\subsection{Quantum Measurement Theory}

As is well known, there's a so called \emph{quantum collapse} in the
quantum measurement theory of the SQM. The reason for this concept
is the superposition of states for single particle. Suppose that the
initial state of a particle is of the form eq.(79), then after a
quantum measurement, the state will collapse into one of the states,
$n$ for example. And according to the SQM, the whole process will be
instantaneous and irreversible. In fact, we have show below eq.(62)
that, the states usually used in SQM are not physical for single
particle but as single-particle ensemble state, and the so called
"collapse" happens only metaphysically or logically as in the
probability theory, not real in nature. However, there are still
some superposition states analyzed in the last subsection, as in
eq.(81),(83) and (84). According to SQM, collapse happens still for
them, but as we will show in this subsection, there is also a
consistent quantum measurement theory for these superposition
states, assuring that nothing unusual will happen.

Now, let's consider a quantum measurement, the famous Stern-Gerlach
experiment for measuring the spins of electrons. However, let's
first replace the non-uniform magnetic field with a uniform
one\footnote{In section V.A., we will propose an operable experiment
to test which interpretation is more proper, the Copenhagen or
ensemble interpretation. }. Let the electrons with spin
$|\uparrow_z>$ travel in this uniform magnetic field, obviously
there's no deflections. With the interaction $B\cdot \hat{n}$,
$\hat{n}=(\sin\theta\cos\varphi,\sin\theta\sin\varphi,\cos\theta)$
for any direction, the final state will be\footnote{$\omega\equiv
|e|B/2m_{e}$. We can see that, the relation between $|\uparrow_z>$
and $|\uparrow_n>$ $|\downarrow_n>$ is changed by the interaction
with the phase terms $e^{\pm i\omega t}$ added in, which are
discussed in the end of last subsection. Then, the state in eq.(85)
is not a superposition state, although it was before the
interaction.}
\begin{eqnarray}
e^{-i\omega t}\cos\frac{\theta}{2}|\uparrow_{n}>+e^{i\omega
t}\sin\frac{\theta}{2}|\downarrow_{n}>,
\end{eqnarray}
a superposition of states. Then, let anther sample of electrons with
spin $|\uparrow_z>$ travel in a non-uniform magnetic field, i.e. the
Stern-Gerlach apparatus. As is known, they will deflect into two
directions, with definite spins $|\uparrow_n>$ and $|\downarrow_n>$
respectively. These two situations are similar physical processes,
but according to the SQM, the conclusions for the state of a single
electron are completely different.

In fact, the above two situations can be described in the unique way
with the field theoretical languages. Here, the field expansion
could be\footnote{The expansion (86) is only a non-relativistic form
because of the constant spinor $u$ in eq.(87). And if the particles
are in states $|\uparrow_n>$ or $|\downarrow_n>$,
$a_{\uparrow_{z}}u_{\uparrow_{z}}+a_{\downarrow_{z}
}u_{\downarrow_{z}}$ is replaced with
$a_{\uparrow_n}u_{\uparrow_n}+a_{\downarrow_n }u_{\downarrow_n}$. }
\begin{eqnarray}
\psi(t,x)=\psi_{\uparrow_{z}}(t,x)+\psi_{\downarrow_{z}}(t,x)=\int
\frac{d^3p}{(2\pi)^3}
[a_{p\uparrow_{z}}u_{\uparrow_{z}}+a_{p\downarrow_{z}
}u_{\downarrow_{z}}]e^{i(p x-E_pt)},
\end{eqnarray}
with spinor representation
\begin{eqnarray}
u_{\uparrow_{z}}={1 \choose 0}\qquad u_{\downarrow_{z}}={0 \choose
1}.
\end{eqnarray}
And the interaction term is
\begin{eqnarray}
H_I=\frac{-e}{2m_e}\int d^3x\psi^\dag(x)\sigma\cdot B(x)\psi(x),
\end{eqnarray}
then the evolution is
\begin{eqnarray}
\exp(-iH_I t)|\uparrow_z>,
\end{eqnarray}
with eq.(85) specific for the uniform  magnetic field\footnote{The
general frequency is of the form $\omega(z)$ with $z$ the direction
of the magnetic field, and when the magnetic field is uniform, the
frequency will be a constant, as in eq.(85). Further,
$\omega(z)t\approx (z-z_0)\partial\omega(z_0)t\approx p_z(z-z_0)$
i.e. the phase in time can be transformed into phase in space,
representing the deflection in the z direction. }. Furthermore, with
$e^{i(p x-E_pt)}$ substituted, eq.(85) can be rewritten
as\footnote{The exact expression is the eq.(215) in section V.A.. }
\begin{eqnarray}
e^{i[px-(E+\omega)t]}\cos\frac{\theta}{2}|\uparrow_{n}>+e^{i[px-(E-\omega)t]}\sin\frac{\theta}{2}|\downarrow_{n}>,
\end{eqnarray}
i.e. the superposition of states with energies $E+\omega$ and
$E-\omega$ for the electrons.

There are two interpretations to eq.(85), one is the SQM version,
assigning a state vector $|\phi(t)>$ describing the evolution of the
state around the sphere; the other one is the QFT version in
eq.(90), in which the time phase terms are parts of the plane waves,
$e^{i[px-(E\pm \omega)t]}$. In the SQM, $|\uparrow_z>$ collapses
irreversibly into $|\uparrow_n>$ or $|\downarrow_n>$, with the
probabilities $|\cos\frac{\theta}{2}|^2$ and
$|\sin\frac{\theta}{2}|^2$ respectively. However, with the
expression (82), under another measurement, $|\uparrow_n>$ may
collapses irreversibly back into $|\uparrow_z>$ again. It appears
that a combination of two irreversible processes could be
reversible. In one word, these statements are a little obscure.
However, with the field theoretical language, whatever the magnetic
field is uniform or not, the descriptions and the conclusions are
definite and unique. From eq.(85) or (90), and according to the
subsection A., we can conclude that, among the sample of electrons
with spin $|\uparrow_z>$, $|\cos\frac{\theta}{2}|^2$ of them whose
states will become $|\uparrow_n>$, while the others will be
$|\downarrow_n>$. Then for one single electron, the probability for
its state to become $|\uparrow_n>$ is $|\cos\frac{\theta}{2}|^2$,
which is just the conclusion of the SQM. Remind that we have assumed
in subsection A. that the state for single particle is definite and
unique, so the processes from $|\uparrow_z>$ to $|\uparrow_n>$, and
$|\uparrow_n>$ back to $|\uparrow_z>$ are all about the \emph{state
transitions} which are unitary without any collapse. In fact, the
essential reason is still that the state in eq.(85) or (90) is an
ensemble state not for single particle. Therefore, we have the
conclusion that the superposition state will change into ensemble
states under the interactions(measurements).

Noting that the evolution in eq.(89) has a similar form as eq.(78),
then one may say that this evolution is the SQM version. In fact, it
is not! Recall the computations of cross sections in QFT, or the
formula of S matrix[2]
\begin{eqnarray}
_{out}<p_1,p_2|k_1,k_2>_{in}\equiv <p_1,p_2|S|k_1,k_2>=\lim_{T\to
\infty}<p_1,p_2|\exp(-iH2T)|k_1,k_2>,
\end{eqnarray}
then in our case, it is about a single particle
\begin{eqnarray}
_{out}<\uparrow_n|\uparrow_z>_{in}=\lim_{T\to
\infty}<\uparrow_n|\exp(-iH2T)|\uparrow_z>,
\end{eqnarray}
similarly for $|\downarrow_n>$. And easily to see, eq.(89) is just
the right hand part of eq.(92), the results are still the transition
amplitudes $<\uparrow_n|\uparrow_z>$ essentially, for the evolution
in eq.(89) is of the form of phase factors as in eq.(85) or (90).

Which description is more proper is now clear, and we can extend the
above discussions to all of the superposition states. As for other
superposition of states, the field theoretical descriptions(or
ensemble concepts) are already proper. According to the example
above, we can give the following new quantum measurement assumption

(4)\emph{Quantum measurement is one kind of $\mathbf{unitary}$ field
interactions. Under the measurement, the states of particles are
unchange if they were just the eigenstates of the measured physical
quantity already, or changed into some of the eigenstates if they
were not before measurement. We could identify the states by some
apparent different macroscopic behaviors, such as deflections in the
above example. What we could obtain is just the probability of
different processes, which is the task of QFT.}\footnote{The changes
of the states of the particles manifest themselves in the change of
the field function, i.e.$\delta\psi(x)$.}

With the modifications above, quantum collapse is completely avoided
for the superposition states. After a measurement, the original
superposition state would change into the ensemble state due to the
arbitrary phase terms resulting from interactions, as in eq.(85).
And the so called collapses occur only metaphysically, not
physically. In addition, the description is the unique
(non-relativistic) QFT, which is a space-time local theory.
Therefore, the so called non-locality in EPR paradox may also be
avoided.

\subsection{EPR Paradox [7]}

We consider the example advocated by Bohm and Aharonov [8]. Let a
pair of spin one-half particles formed in the singlet spin
state\footnote{Here, $|\uparrow>$ can be at any direction, because
of the total spin is zero for $|\Psi>_{AB}$, so we can treat it as
$|\uparrow_z>$. This is only for this spin singlet, not for other
Bell states, for example,
$\frac{1}{\sqrt{2}}(|\uparrow\downarrow>_{AB}+|\downarrow\uparrow>_{AB})$.}
\begin{eqnarray}
|\Psi>_{AB}=\frac{1}{\sqrt{2}}(|\uparrow\downarrow>_{AB}-|\downarrow\uparrow>_{AB}),
\end{eqnarray}
move freely in opposite directions. Assume that we make a
measurement for the particle A, then according to the SQM, there
will be quantum collapse, if A is found to be at $|\uparrow>_A$, the
state of B will be collapsed into $|\downarrow>_B$. This collapse is
instantaneous, so that we can construct two events in space-time,
which are separated by a space-like interval, one is the measurement
for A, the other is the one for B. Then the relativity causality and
locality are violated.

Obviously, the violation of causality and locality is also owing to
the quantum collapse, which, as we have described above, could be
completely avoided in the field theoretical language. In fact, as we
have analyzed in subsection B., the state in eq.(93) is an example
of superposition states as in eq.(84),i.e. the coupling of angular
moentua. In addition to the state in eq.(93), there are another
three, one of which is
\begin{eqnarray}
|\Phi>_{AB}=\frac{1}{\sqrt{2}}(|\uparrow\downarrow>_{AB}+|\downarrow\uparrow>_{AB}).
\end{eqnarray}
With these two states, we can express $|\uparrow\downarrow>_{AB}$ as
follows
\begin{eqnarray}
|\uparrow\downarrow>_{AB}=\frac{1}{\sqrt{2}}(|\Phi>_{AB}+|\Psi>_{AB}),
\end{eqnarray}
similarly for the other one. Then according to the quantum
measurement in SQM, with a special quantum measurement, we could
obtain the so called quantum entangled states, just like the case in
the last subsection for the single spin states. These are also
obscure, so we need the field theoretical language.

All the things are already studied in the last subsection. What we
need are eqs.(88) and (89) for the measurements. First, we make a
measurement for A, after that, the state in eq.(93) will become
\begin{eqnarray}
\frac{1}{\sqrt{2}}(e^{-i\omega_A
t}|\uparrow_n\downarrow_n>_{AB}-e^{i\omega_A
t}|\downarrow_n\uparrow_n>_{AB}).
\end{eqnarray}
Then for B, we have\footnote{Notice that eqs.(96) and (97) are of
the forms of eq.(64) for two particles within two states.}
\begin{eqnarray}
\frac{1}{\sqrt{2}}(e^{-i\omega_A t}e^{i\omega_B
t'}|\uparrow_n\downarrow_n>_{AB}-e^{i\omega_A t}e^{-i\omega_B
t'}|\downarrow_n\uparrow_n>_{AB}).
\end{eqnarray}
Of course, we have assumed that the directions of the magnetic field
are the same for both the measurements, and for different
directions, the expression will be complicated.

From eqs.(96) and (97), we can claim that:

(1)The total spin is not conserved during the measurements, which is
easily to understand, because the interactions are spin dependent,
there are angular momentum exchanges between the particles and the
magnetic field (or the photons). Only if the directions of the
magnetic field were the same for both the measurements, the spin in
that direction would be conserved.

(2)The measurements for A and B are independent, because they are
field interactions, so we cannot construct two events which violate
the relativity causality and locality.

(3)According to our ideas, the states in eqs.(96) and (97) are not
superposition states because of the arbitrary phase terms, and from
eq.(97), we could obtain that, the probability for the transition
from $|\Psi>_{AB}$ to $|\uparrow_n\downarrow_n>_{AB}$ is one-half.

Here is one important note about eq.(96), from which one may say
that, the state of B is changed instantaneously. In fact, the state
of B is still the initial one $|\uparrow>_B$ or $|\downarrow>_B$,
the expression for eq.(96) is just for convenient mathematically. In
fact, with the following representation of the state in eq.(93)
\begin{eqnarray}
|\Psi>_{AB}=\frac{1}{\sqrt{2}}(|\uparrow_z\downarrow_z>_{AB}-|\downarrow_z\uparrow_z>_{AB}),
\end{eqnarray}
the original eq.(96) will be
\begin{eqnarray}
\frac{1}{\sqrt{2}}[e^{-i\omega_A
t}|\uparrow_n>_A(e^{-i\frac{\varphi}{2}}\sin\frac{\theta}{2}|\uparrow_z>_{B}-e^{i\frac{\varphi}{2}}\cos\frac{\theta}{2}|\downarrow_z>_{B})\nonumber\\
-e^{i\omega_A
t}|\downarrow_n>_A(e^{-i\frac{\varphi}{2}}\cos\frac{\theta}{2}|\uparrow_z>_{B}+e^{i\frac{\varphi}{2}}\sin\frac{\theta}{2}|\downarrow_z>_{B})].
\end{eqnarray}
Although the parts for B can be rewritten compactly as in eq.(96),
we could obtain the physical results easily and consistently from
eq.(99), for instance, with a measurement only for A, we should sum
over all the possibilities about B. For example, the probability for
A to be at state $|\uparrow_n>_A$ is
\begin{eqnarray}
P_{|\uparrow_n\uparrow_z>}+P_{|\uparrow_n\downarrow_z>}=\frac{1}{2}\times
(|\cos\frac{\theta}{2}|^2+|\sin\frac{\theta}{2}|^2)=\frac{1}{2},
\end{eqnarray}
the same for the other one. If the magnetic field for measuring B is
different from that of A, eq.(99) will be a good starting point.

All the above descriptions can be written in more field theoretical
forms. For example, the state in eq.(93) is
\begin{eqnarray}
\Psi_{AB}(x_1,x_2)=\psi_\uparrow(x_1)_A\psi_\downarrow(x_2)_B-\psi_\downarrow(x_1)_A\psi_\uparrow(x_2)_B,
\end{eqnarray}
and the variation due to the measurement for A is (first order)
\begin{eqnarray}
\delta\Psi_{AB}(x_1,x_2)=\delta\psi_\uparrow(x_1)_A\psi_\downarrow(x_2)_B-\delta\psi_\downarrow(x_1)_A\psi_\uparrow(x_2)_B,
\end{eqnarray}
or
\begin{eqnarray}
\delta\Psi_{AB}=iT[\frac{-e}{2m_e}\int
d^3yB(y)\psi^\dag_A(\sigma_A\cdot \hat{n})\psi_A(y),\Psi_{AB}].
\end{eqnarray}
In Bell's paper [9], there is a correlation function
\begin{eqnarray}
P(a,b)=_{AB}<\Psi|(\sigma _A\cdot a)(\sigma _B\cdot b)|\Psi>_{AB},
\end{eqnarray}
with eq.(49), the corresponding field theoretical expression is
\begin{eqnarray}
\frac{_{AB}<\Psi|\int d^3x_1\psi^\dag_A(\sigma _A\cdot
a)\psi_A(x_1)\int d^3x_2\psi^\dag_B(\sigma _B\cdot
b)\psi_B(x_2)|\Psi>_{AB}}{_{AB}<\Psi|\int
d^3x_1\psi^\dag_A\psi_A(x_1)\int
d^3x_2\psi^\dag_B\psi_B(x_2)|\Psi>_{AB}}.
\end{eqnarray}
or the one similar to eq.(59)\footnote{In other words, the particle
pair should be considered as one system, just like the
single-particle case with states $|\uparrow_n\downarrow_n>_{AB}$,
$|\downarrow_n\uparrow_n>_{AB}$, etc. }.

There is still a problem in the analysis above, the field function
in eq.(101) is not identical to the following expression
\begin{eqnarray}
\psi_{\uparrow_n}(x_1)_A\psi_{\downarrow_n}(x_2)_B-\psi_{\downarrow_n}(x_1)_A\psi_{\uparrow_n}(x_2)_B.
\end{eqnarray}
The difference between them can be canceled in a artificial way, and
the exact expression should be relativistic. After all, the EPR
paradox is solved, it's just a misunderstanding.

Furthermore, with our ideas, the so called quantum entanglement is
also not real in nature, and among the four Bell states in SQM,
eqs.(93) and (94) are physical states, while the other two
\begin{eqnarray}
\frac{1}{\sqrt{2}}(|\uparrow\uparrow>_{AB}\pm
|\downarrow\downarrow>_{AB}),
\end{eqnarray}
are only superposition of states $|\uparrow\uparrow>_{AB}$ and
$|\downarrow\downarrow>_{AB}$. And the example shown in the original
EPR paper can also be solved since it just involves the
transformations between the lab frame and center-of-mass frame, as
described below eq.(83) in subsection B..

\subsection{Double-slit Interference Experiment}

In QM, the most famous experiment must be the double-slit
interference experiment which is believed to contain the essential
features of QM. In this subsection, we will study this experiment in
details by using the concepts described previously, and obtain the
required particle number distribution. The experiment is sketched in
Fig.1. From the original point $O$, there will be a beam of
particles moving to the double-slit screen, if the two slits are
both open, then we will obtain interference fringes on the receiving
screen. However, if we control the slits so that they are open not
at the same time, then the interference fringes would disappear.

\begin{figure}
\setlength{\unitlength}{1mm} \centering
\includegraphics[width=3.0in]{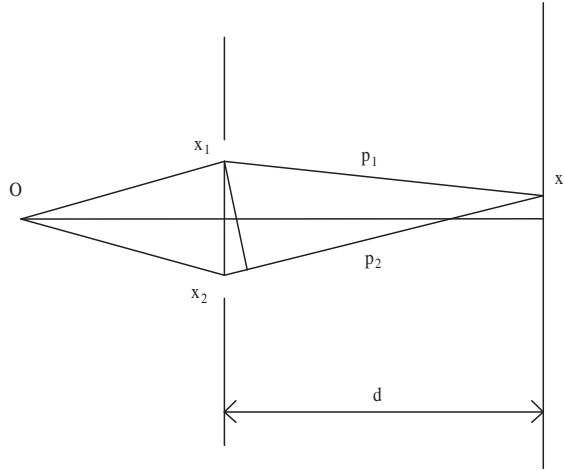}
\caption[]{Double-slit interference experiment.}\label{fig1}
\end{figure}

There is a rough QM description with the use of the wave properties
of quantized particles as follows. For an arbitrary point $x$ on the
receiving screen, there will be two waves $\psi_1(x)$ and
$\psi_2(x)$ coming from the two slits respectively, then the total
wave will be
\begin{eqnarray}
\psi(x)=\psi_1(x)+\psi_2(x),
\end{eqnarray}
and according to QM, we should compute the probability
$|\psi(x)|^2$, then there will be crossing interference terms. In
fact, the interference can also be considered to be from the path
difference as sketched in Fig.1, from the point of view of pure
classical waves, such as the light waves.

Now, let's give a field theoretical description. Eq.(108) is still
proper, with the wave functions interpreted as fields. The path
difference in wave theory is actually phase difference which can be
resulting from the interactions of the particles with the
double-slit screen. The interaction can be considered to be elastic
collision, and under the interaction, the energies of the particles
are unchange while the momenta are changed. We can describe this
process with the following form in first order
\begin{eqnarray}
\psi(x,x_0)=\int d^4y\int d^4zK(x,x_0;y,y_0)
V(y)K(y,y_0;z,z_0)\psi(z,z_0),
\end{eqnarray}
with the propagator defined as in eq.(34)
\begin{eqnarray}
K(x,x_0;y,y_0)= <0|\psi(x,x_0)\psi^\dag(y,y_0)|0>.
\end{eqnarray}
We assume the following interactions\footnote{Notice that if we had
infinite slits on the screen, then the interactions would be
$\delta^3(y-x_1)+\delta^3(y-x_2)+\cdots=\int d^3x\delta^3(y-x)=1$,
and eq.(109) would be just the combination of two propagators.}
\begin{eqnarray}
V(y)=\delta^3(y-x_1)+\delta^3(y-x_2),
\end{eqnarray}
and after simple computations, we will obtain the dependence of the
field function on the $x_1$ and $x_2$. In fact, we can obtain them
in a much simpler way, note that the two slits are actually two
sources as in eq.(111), and the field equation will be that of
propagator with the source terms. In one word, we can use the
propagator in eq.(110) as a basis. Since the propagator is for the
free particle, the field function in the interval $[x,x+dx]$ will
be\footnote{The momenta of the particles will roughly be constant in
this case. }
\begin{eqnarray}
\psi(x)=a_{p_1}e^{ip_1(x-x_1)}+a_{p_2}e^{ip_2(x-x_2)} ,
\end{eqnarray}
then the particle number density $N(x)=\psi^\dag(x)\psi(x)$ is
\begin{eqnarray}
N(x)=a^{\dag}_{p_1}a_{p_1}+a^{\dag}_{p_2}a_{p_2}+e^{i\alpha(x,x_1,x_2)}a^{\dag}_{p_1}a_{p_2}+e^{-i\alpha(x,x_1,x_2)}a^{\dag}_{p_2}a_{p_1},
\end{eqnarray}
where we have collected the phase terms in a compact from, and
easily to see they will cause the interference.

The next task is to find out the state of ensemble for the two
momentum states, for example the Bosonic case in eq.(64) with $k=2$
\begin{eqnarray}
\sum^{n}
_{j=0}\sqrt{\frac{P_{j}}{j!(n-j)!}}(a_{p_1}^\dag)^{j}(a_{p_2}^\dag)^{n-j}|0>,
\end{eqnarray}
where the arbitrary phase terms are already absorbed into eq.(113).
We need the mean value of particle number density $<N(x)>$ with the
state in eq.(114) substituted. For the diagonal term, the result is
just $n$, while for the off-diagonal terms we will have
\begin{eqnarray}
\sum^{n} _{j=0}2\sqrt{P_{j}P_{j+1}}\sqrt{(j+1)(n-j)}\cos\alpha.
\end{eqnarray}
For simplicity, we assume that the probability is
\begin{eqnarray}
P_{j}=\left(\frac{1}{2}\right)^n\frac{n!}{j!(n-j)!},
\end{eqnarray}
which is related to the binomial coefficients, and substituting it
into eq.(115), we then have
\begin{eqnarray}
n\cos\alpha,
\end{eqnarray}
which is the interference term! Therefore, the total particle number
distribution is
\begin{eqnarray}
V<N(x)>=n(1+\cos\alpha(x,x_1,x_2)),
\end{eqnarray}
with $V$ the space volume, the normalization of plane wave which is
ignored for convenient previously. Obviously, eq.(118) is also
proper for the single particle ensemble state, for instance,
$(|p_1>+|p_2>)/\sqrt{2}$, with single particle in the whole space,
i.e. $n=1$\footnote{This is also for the Fermionic case with the
Pauli exclusion principle, ignoring the spins.}. In this sense, we
could also obtain the above special result in a simpler way, by
noting that the above n-particle ensemble with probabilities in
eq.(116) is in fact made up of single particle ensemble as noted in
the end of subsection A.. Now, we rewrite the filed in the ordinary
form
\begin{eqnarray}
\psi(x)=a_{p_1}e^{ip_1x}+a_{p_2}e^{ip_2x}+\cdots ,
\end{eqnarray}
and in order to obtain the exact interference term, we should have
the following single particle ensemble state
\begin{eqnarray}
|>=\frac{1}{\sqrt{2}}(e^{-ip_1x_1}|p_1>+e^{-ip_2x_2}|p_2>) ,
\end{eqnarray}
with the respective phase terms added. Then the $<N(x)>$ is
\begin{eqnarray}
V<N(x)>=\frac{1}{2}(2+e^{i\alpha(x,x_1,x_2)}+e^{-i\alpha(x,x_1,x_2)})=1+\cos\alpha(x,x_1,x_2),
\end{eqnarray}
which is just the $n=1$ case of eq.(118)! To obtain the general
formula eq.(118), we construct the $n$-particle ensemble out of the
state in eq.(120), obtaining the state in eq.(114) with the phase
terms already absorbed into eq.(113)\footnote{Notice that
$<p_1|e^{ip_1x_1}\psi^\dag(x)\psi(x)e^{-ip_1x_1}|p_1>=<p_1|e^{iPx_1}\psi^\dag(x)\psi(x)e^{-iPx_1}|p_1>=<p_1|\psi^\dag(x-x_1)\psi(x-x_1)|p_1>$.
}, and the probability condition in eq.(116). Now, if we control the
slits so that they are open not at the same time, so that the source
of each particle at the interval $[x,x+dx]$ are definite, in other
words, the distribution $[n_j]$ in eq.(64) is determined in this
case, then the state may be, for example
\begin{eqnarray}
\frac{1}{\sqrt{(n_1)!(n_2)!}}(a_{p_1}^\dag)^{n_1}(a_{p_2}^\dag)^{n_2}|0>,\qquad
n_1+n_2=n,
\end{eqnarray}
i.e. a measurement which causes a collapse of the ensemble state in
eq.(114), then the off-diagonal terms in $<N(x)>$ disappear, so do
the interferences.

Notice that, this field theoretical description includes the QM
description in eq.(108), by using the wave functions defined in
eq.(77) with the single particle ensemble state in eq.(120). Even it
can include the classical field case with the coherent state defined
in eqs.(228) and (238).

\section{Relativistic Extensions}

In section II., we have studied in details the non-relativistic
(quantum)Schr\"odinger field theory which can be considered to be
much more fundamental than the QM, for all the QM can be derived
from this field theory. With this new approach to QM, modifications
to the SQM is developed in section III., where the ensemble
interpretation is realized by treating most of the states in QM as
ensemble states, such as the ones in eqs.(62) and (64), while the
rest as a class of superposition state as shown in eqs.(81)(83) and
(84). In this section, we will extend these concepts to the
relativistic QFT, indicating that fields are the fundamental
elements of the physical world and QFT is the unique consistent
theory by now.

As demonstrated in section II., the most important elements are the
ordinary QM physical operators(about particles) made up of fields,
such as the operators in eqs.(5)-(8), and the corresponding
eigenstates. Since the energy and momentum operators can be obtained
from transformations of the action under the space-time translation,
what we need are only the rest two, the particle number and position
operators. Notice from eqs.(7) and (8) that, the position operator
may be considered to be followed by substituting a space coordinate
into the formula of particle number operator, so the only thing we
need is to find out the general rule for the particle number
operator. Easily to see, there is a gauge symmetry of the action (1)
with the fields transform as, in quantized form
\begin{eqnarray}
\psi\rightarrow e^{-i\alpha}\psi,\qquad \psi^\dag\rightarrow
e^{i\alpha}\psi^\dag,
\end{eqnarray}
then there is a physical quantity corresponding to this symmetry,
just like the ordinary $U(1)$ gauge, and obviously this physical
quantity is the particle number!

However, as is well known, in the ordinary relativistic QFT formula,
it seems to be impossible to impose this symmetry, for the particle
and anti-particle fields are written in some special combined form,
for example a free charged scalar field\footnote{In this section and
below, we use the bold face letters to denote the vector form of the
three space dimensional coordinates and momenta.}
\begin{eqnarray}
\phi(\mathbf{x})=\phi_1(\mathbf{x})+\phi^\dag_2(\mathbf{x})=\int
\frac{d^3\mathbf{p}}{(2\pi)^3}\frac{1}{\sqrt{2E_{\mathbf{p}}}}
(a(\mathbf{p})e^{i\mathbf{px}}+b^\dag(\mathbf{p})e^{-i\mathbf{px}}),
\end{eqnarray}
to guarantee the communicative relation
$[\phi(\mathbf{x}),\phi^\dag(\mathbf{y})]=0$. But if we treat the
fields $\phi_1(\mathbf{x})$ and $\phi_2(\mathbf{x})$ as independent,
we could construct another combined field in the following form
\begin{eqnarray}
\phi'(\mathbf{x})=i(\phi_1(\mathbf{x})-\phi^\dag_2(\mathbf{x})),
\end{eqnarray}
which obviously also satisfies the corresponding communicative
relation $[\phi'(\mathbf{x}),\phi'^\dag(\mathbf{y})]=0$. Now, we
write down the filed action(in quantized form)
\begin{eqnarray}
S=\int d^4x
(\partial_\mu\phi^\dag\partial^\mu\phi-m^2\phi^\dag\phi),
\end{eqnarray}
then substituting the two fields in eqs.(124) and (125) into it, we
have
\begin{eqnarray}
S_\phi=\int d^4x
[(\partial_\mu\phi_1^\dag\partial^\mu\phi_1-m^2\phi_1^\dag\phi_1)+(\partial_\mu\phi_2\partial^\mu\phi_2^\dag-m^2\phi_2\phi_2^\dag)\nonumber\\
+(\partial_\mu\phi_1^\dag\partial^\mu\phi_2^\dag-m^2\phi_1^\dag\phi_2^\dag)+(\partial_\mu\phi_2\partial^\mu\phi_1-m^2\phi_2\phi_1)],
\end{eqnarray}
and
\begin{eqnarray}
S_{\phi'}=\int d^4x
[(\partial_\mu\phi_1^\dag\partial^\mu\phi_1-m^2\phi_1^\dag\phi_1)+(\partial_\mu\phi_2\partial^\mu\phi_2^\dag-m^2\phi_2\phi_2^\dag)\nonumber\\
-(\partial_\mu\phi_1^\dag\partial^\mu\phi_2^\dag-m^2\phi_1^\dag\phi_2^\dag)-(\partial_\mu\phi_2\partial^\mu\phi_1-m^2\phi_2\phi_1)].
\end{eqnarray}
Easily to see, the combination $\frac{1}{2}(S_\phi+S_{\phi'})$ is
what we need
\begin{eqnarray}
S'=\int d^4x
[(\partial_\mu\phi_1^\dag\partial^\mu\phi_1-m^2\phi_1^\dag\phi_1)+(\partial_\mu\phi_2\partial^\mu\phi_2^\dag-m^2\phi_2\phi_2^\dag),
\end{eqnarray}
from which the fields $\phi_1$ and $\phi_2$ are independent from
each other, thus we could consider separately the particle field and
anti-particle field. Let's consider the particle field $\phi_1$ and
its action $S_{\phi_1}$, the canonical momenta are\footnote{The
communicative relations among the four quantities, $\phi_1$,
$\phi^\dag_1$, $\pi_{\phi_1}$ and $\pi_{\phi^\dag_1}$ can be
computed by using the expansion in eq.(124), which are different
from the relations among $\phi$, $\phi^\dag$, $\pi_{\phi}$ and
$\pi_{\phi^\dag}$.}
\begin{eqnarray}
\pi_{\phi_1}=\dot{\phi^\dag_1},\qquad
\pi_{\phi^\dag_1}=\dot{\phi_1},
\end{eqnarray}
then the Hamiltonian density is
\begin{eqnarray}
\mathcal
{H}_{\phi_1}=\pi_{\phi_1}\dot{\phi_1}+\pi_{\phi^\dag_1}\dot{\phi^\dag_1}-\mathcal
{L}_{\phi_1}=\pi_{\phi^\dag_1}\pi_{\phi_1}+\nabla\phi^\dag_1\nabla\phi_1+m^2\phi^\dag_1\phi_1,
\end{eqnarray}
after substituting the expansion of $\phi_1$ in eq.(124), we will
have the energy of the field
\begin{eqnarray}
H_{\phi_1}=\int d^3\mathbf{x}\mathcal {H}_{\phi_1}=\int
\frac{d^3\mathbf{p}}{(2\pi)^3}\frac{E_{\mathbf{p}}}{2}(a(\mathbf{p})a^\dag(\mathbf{p})+a^\dag(\mathbf{p})a(\mathbf{p})),
\end{eqnarray}
similarly, the momentum of the field is
\begin{eqnarray}
\mathbf{P}_{\phi_1}=-\int
d^3\mathbf{x}(\pi_{\phi_1}\nabla\phi_1+\pi_{\phi_1^\dag}\nabla\phi^\dag_1)=\int
\frac{d^3\mathbf{p}}{(2\pi)^3}\frac{\mathbf{p}}{2}(a(\mathbf{p})a^\dag(\mathbf{p})+a^\dag(\mathbf{p})a(\mathbf{p})).
\end{eqnarray}
Since the action $S_{\phi_1}$ has a similar form as the action in
eq.(1), there is also a gauge transformation of $\phi_1$
\begin{eqnarray}
\phi_1\rightarrow e^{-i\alpha}\phi_1,\phi_1^\dag\rightarrow
e^{i\alpha}\phi_1^\dag,
\end{eqnarray}
from which we obtain a physical quantity, i.e. the particle number
\begin{eqnarray}
N_{\phi_1}=-i\int
d^3\mathbf{x}(\pi_{\phi_1}\phi_1-\pi_{\phi_1^\dag}\phi^\dag_1)=\frac{1}{2}\int
\frac{d^3\mathbf{p}}{(2\pi)^3}(a(\mathbf{p})a^\dag(\mathbf{p})+a^\dag(\mathbf{p})a(\mathbf{p})).
\end{eqnarray}
Thus, a possible position operator can be defined as
\begin{eqnarray}
\mathbf{X}_{\phi_1}=-i\int
d^3\mathbf{x}\mathbf{x}(\pi_{\phi_1}\phi_1-\pi_{\phi_1^\dag}\phi^\dag_1),
\end{eqnarray}
and after a simple computation we have
\begin{eqnarray}
\mathbf{X}_{\phi_1}=i\int
\frac{d^3\mathbf{p}}{(2\pi)^3}[\frac{1}{2}a^\dag(\mathbf{p})\partial_{\mathbf{p}}a(\mathbf{p})+\frac{1}{2}\partial_{\mathbf{p}}a(\mathbf{p})a^\dag(\mathbf{p})
\nonumber\\+a^\dag(\mathbf{p})a(\mathbf{p})\sqrt{\frac{E_{\mathbf{p}}}{2}}\partial_{\mathbf{p}}(\frac{1}{\sqrt{2E_{\mathbf{p}}}})
+a(\mathbf{p})a^\dag(\mathbf{p})\partial_{\mathbf{p}}(\sqrt{\frac{E_{\mathbf{p}}}{2}})\frac{1}{\sqrt{2E_{\mathbf{p}}}}],
\end{eqnarray}
where the last two terms will vanish, using the communicative
relation
$[a(\mathbf{p}),a^\dag(\mathbf{p})]=(2\pi)^3\delta^3(\mathbf{0})$
and $\int
d^3\mathbf{p}\mathbf{p}F(\mathbf{p})=0(F(\mathbf{p})=F(-\mathbf{p}))$.
Then the position operator become
\begin{eqnarray}
\mathbf{X}_{\phi_1}=i\int
\frac{d^3\mathbf{p}}{(2\pi)^3}a^\dag(\mathbf{p})\partial_{\mathbf{p}}a(\mathbf{p}),
\end{eqnarray}
which is just the eq.(17)! Up to the orders of  the creators and
annihilators, we have obtained all the required physical operators
for particles.

The next task is to find out the eigenstates of these operators as
in section II. The momentum state is already defined well in QFT[2]
\begin{eqnarray}
|\mathbf{p}>=\sqrt{2E_{\mathbf{p}}}a^\dag(\mathbf{p})|0>,
\end{eqnarray}
with normalization
\begin{eqnarray}
<\mathbf{p}|\mathbf{q}>=2E_{\mathbf{p}}(2\pi)^3\delta^3(\mathbf{p-q}).
\end{eqnarray}
Noting that this definition has a good Lorentz transformation
property, for $E_{\mathbf{p}}\delta^3(\mathbf{p-q})$ is Lorentz
invariant. However, it seems to be impossible for this property to
be imposed on space-time\footnote{For a real particle, its energy
and momentum satisfy the condition
$E^2_{\mathbf{p}}=\mathbf{p}^2+m^2$, but there is not a general
relation between space and time coordinates of the particles. },
that is we can only have the following normalization
\begin{eqnarray}
<\mathbf{x}|\mathbf{y}>=\delta^3(\mathbf{x-y}).
\end{eqnarray}
One may simply thought that $\phi_1^\dag(\mathbf{x})|0>$ be the
required position eigenstate, but the factor
$1/\sqrt{2E_{\mathbf{p}}}$ in the field expansion would make the
problems more complicated. Recalling the state in eq.(20) and the
filed expansion in eq.(14), we can define a new "field"
\begin{eqnarray}
\psiup_{\phi_1}(\mathbf{x})=\int \frac{d^3\mathbf{p}}{(2\pi)^3}
a(\mathbf{p})e^{i\mathbf{px}},
\end{eqnarray}
with the communicative relation
\begin{eqnarray}
[\psiup_{\phi_1}(\mathbf{x}),\psiup_{\phi_1}^\dag(\mathbf{y})]=\delta^3(\mathbf{x-y}),
\end{eqnarray}
then the position eigenstate $|\mathbf{x}>$ can be defined as
\begin{eqnarray}
|\mathbf{x}>\equiv \psiup_{\phi_1}^\dag(\mathbf{x})|0>,
\end{eqnarray}
with the required normalization in eq.(141) by using eq.(143).

Here, let's have a look at the "field" $\psiup_{\phi_1}(\mathbf{x})$
defined in eq.(142), obviously it is not a well defined filed,
because its Lorentz transformation is obscure. But it's indeed
useful for us, with it, we could redefine all the above physical
operators as
\begin{eqnarray}
H_{\phi_1}=\int d^3\mathbf{x}\psiup_{\phi_1}^\dag(\mathbf{x})\hat
{H}\psiup_{\phi_1}(\mathbf{x})=\int
\frac{d^3\mathbf{p}}{(2\pi)^3}E_{\mathbf{p}}a^\dag(\mathbf{p})a(\mathbf{p})\\
\mathbf{P}_{\phi_1}=\int
d^3\mathbf{x}\psiup_{\phi_1}^\dag(\mathbf{x})\hat
{\mathbf{p}}\psiup_{\phi_1}(\mathbf{x})=\int
\frac{d^3\mathbf{p}}{(2\pi)^3}\mathbf{p}a^\dag(\mathbf{p})a(\mathbf{p})\\
N_{\phi_1}=\int
d^3\mathbf{x}\psiup_{\phi_1}^\dag(\mathbf{x})\psiup_{\phi_1}(\mathbf{x})=\int
\frac{d^3\mathbf{p}}{(2\pi)^3}a^\dag(\mathbf{p})a(\mathbf{p})\\
\mathbf{X}_{\phi_1}=\int
d^3\mathbf{x}\psiup_{\phi_1}^\dag(\mathbf{x})\hat
{\mathbf{x}}\psiup_{\phi_1}(\mathbf{x})=i\int
\frac{d^3\mathbf{p}}{(2\pi)^3}a^\dag(\mathbf{p})\partial_{\mathbf{p}}a(\mathbf{p}),
\end{eqnarray}
with the single particle operators defined as
\begin{eqnarray}
\hat {H}=\sqrt{{\hat {\mathbf{p}}}^2+m^2},\qquad \hat
{\mathbf{p}}=-i\mathbf{\nabla}, \qquad \hat {\mathbf{x}}=\mathbf{x},
\end{eqnarray}
which are the familiar operators of QM in the relativistic forms.

Now, let's study the Lorentz transformation $\Lambda$, which will be
implemented as some unitary operator $U(\Lambda)$, and for the
momentum state in eq.(139), we have[2]
\begin{eqnarray}
U(\Lambda)|\mathbf{p}>=|\Lambda\mathbf{p}>,
\\U(\Lambda)a^\dag(\mathbf{p})U^{-1}(\Lambda)=\sqrt{\frac{E_{\Lambda\mathbf{p}}}{E_{\mathbf{p}}}}a^\dag(\Lambda\mathbf{p}).
\end{eqnarray}
Considering a boost in the 3-direction $p'_3=\gamma(p_3+\beta
E),E'=\gamma(E+\beta p_3)$, the operators in eqs.(145)-(148) should
have the ordinary transformation properties, noting that $\int
\frac{d^3\mathbf{p}}{(2\pi)^3}\frac{1}{2E_{\mathbf{p}}}$ is Lorentz
invariant, for example, the energy and momentum operators transform
as
\begin{eqnarray}
UHU^{-1}=\gamma(H-\beta P_3),UP_3U^{-1}=\gamma(P_3-\beta H),
\end{eqnarray}
while the particle number operator is Lorentz invariant. The last
one is the transformation of the position operator $UX_3U^{-1}$,
where the eq.(151) makes the problem complicated. To solve it, we
introduce operators $\tilde{a^\dag}$ and $\tilde{a}$
\begin{eqnarray}
\tilde{a}^\dag(\mathbf{p})=\sqrt{2E_{\mathbf{p}}}a^\dag(\mathbf{p}),\qquad
\tilde{a}(\mathbf{p})=\sqrt{2E_{\mathbf{p}}}a(\mathbf{p}),
\end{eqnarray}
with the communicative relation
\begin{eqnarray}
[\tilde{a}(\mathbf{p}),\tilde{a}^\dag(\mathbf{q})]=2E_{\mathbf{p}}(2\pi)^3\delta^3(\mathbf{p}-\mathbf{q}),
\end{eqnarray}
then
\begin{eqnarray}
\mathbf{X}=\frac{i}{2}\int
\frac{d^3\mathbf{p}}{(2\pi)^3}[a^\dag(\mathbf{p})\partial_{\mathbf{p}}a(\mathbf{p})-\partial_{\mathbf{p}}a^\dag(\mathbf{p})a(\mathbf{p})]
\nonumber\\=\frac{i}{2}\int
\frac{d^3\mathbf{p}}{(2\pi)^3}\frac{1}{2E_{\mathbf{p}}}[\tilde{a}^\dag(\mathbf{p})\partial_{\mathbf{p}}\tilde{a}(\mathbf{p})-\partial_{\mathbf{p}}\tilde{a}^\dag(\mathbf{p})\tilde{a}(\mathbf{p})].
\end{eqnarray}
Thus we have
\begin{eqnarray}
UX_3U^{-1}=\frac{i}{2}\int
\frac{d^3\mathbf{p'}}{(2\pi)^3}\frac{1}{2E_{\mathbf{p'}}}[\tilde{a}^\dag(\mathbf{p'})\gamma(\partial_{p'_3}+\beta\partial_{E'})\tilde{a}(\mathbf{p'})
\nonumber\\-\gamma(\partial_{p'_3}+\beta\partial_{E'})\tilde{a}^\dag(\mathbf{p'})\tilde{a}(\mathbf{p'})]
=\gamma(X_3-\beta T),
\end{eqnarray}
where we have use the transformation property
\begin{eqnarray}
U(\Lambda)\tilde{a}^\dag(\mathbf{p})U^{-1}(\Lambda)=\tilde{a}^\dag(\Lambda\mathbf{p}),
\end{eqnarray}
and defined a time operator
\begin{eqnarray}
T=-\frac{i}{2}\int
\frac{d^3\mathbf{p}}{(2\pi)^3}\frac{1}{2E_{\mathbf{p}}}[\tilde{a}^\dag(\mathbf{p})\partial_{E}\tilde{a}(\mathbf{p})
-\partial_{E}\tilde{a}^\dag(\mathbf{p})\tilde{a}(\mathbf{p})]\nonumber\\=-\frac{i}{2}\int
\frac{d^3\mathbf{p}}{(2\pi)^3}[a^\dag(\mathbf{p})\partial_{E}a(\mathbf{p})
-\partial_{E}a^\dag(\mathbf{p})a(\mathbf{p})],(\partial_{E}=\frac{\partial\mathbf{p}}{\partial
E}\cdot\partial_{\mathbf{p}}),
\end{eqnarray}
which transforms as
\begin{eqnarray}
UTU^{-1}=-\frac{i}{2}\int
\frac{d^3\mathbf{p'}}{(2\pi)^3}\frac{1}{2E_{\mathbf{p'}}}[\tilde{a}^\dag(\mathbf{p'})\gamma(\partial_{E'}+\beta\partial_{p'_3})\tilde{a}(\mathbf{p'})
\nonumber\\-\gamma(\partial_{E'}+\beta\partial_{p'_3})\tilde{a}^\dag(\mathbf{p'})\tilde{a}(\mathbf{p'})]=\gamma(T-\beta
X_3),
\end{eqnarray}
consistent with the above transformations, in other words,
$P_\mu=(H,-\mathbf{P})$ and $X^\mu=(T,\mathbf{X})$ are two 4-vector
operators. Further, we have the following communicative relation
between $H$ and $T$
\begin{eqnarray}
[T,H]=-i\int
\frac{d^3\mathbf{p}}{(2\pi)^3}a^\dag(\mathbf{p})a(\mathbf{p})=-i N,
\end{eqnarray}
which combined with eq.(13) makes up the diagonal terms
$-i\eta^{\mu\nu}N$ of the communicative relation $[X^\mu,P^\nu]$,
leaving the terms $[\mathbf{X},H]$ and $[T,\mathbf{P}]$.

Now, we have to find out the eigenstate of time operator $T$, and
after a few tedious calculations, we have
\begin{eqnarray}
[\mathbf{X},T]=\frac{1}{2}\int
\frac{d^3\mathbf{p}}{(2\pi)^3}\{a^\dag(\mathbf{p})[\partial_{\mathbf{p}},\partial_{E_{\mathbf{p}}}]a(\mathbf{p})
-[\partial_{\mathbf{p}},\partial_{E_{\mathbf{p}}}]a^\dag(\mathbf{p})a(\mathbf{p})\},
\end{eqnarray}
and since
\begin{eqnarray}
[\partial_{\mathbf{p}_i},\partial_{E_{\mathbf{p}}}]=\sum_{j}\partial_{\mathbf{p}_i}(\frac{\partial\mathbf{p}_j}{\partial
E_{\mathbf{p}}})\partial_{\mathbf{p}_j},
\end{eqnarray}
thus the position and time operators have different eigenstates. To
find it, let's propose the following form
\begin{eqnarray}
|t>=\Phi^\dag(t,\mathbf{x}=\mathbf{0})|0>=\int
\frac{d^3\mathbf{p}}{(2\pi)^3}\phi(\mathbf{p})e^{iE_{\mathbf{p}}t}a^\dag(\mathbf{p})|0>,
\end{eqnarray}
with an undetermined factor $\phi(\mathbf{p})$, and this state
should satisfy $T|t>=t|t>$, or the communicative relation
\begin{eqnarray}
[T,\Phi^\dag(t,\mathbf{x}=\mathbf{0})]=t\Phi^\dag(t,\mathbf{x}=\mathbf{0}),
\end{eqnarray}
from which we get a differential equation about $\phi(\mathbf{p})$
\begin{eqnarray}
2\partial_{E_{\mathbf{p}}}\phi(\mathbf{p})+\sum_{i}\partial_{\mathbf{p}_i}(\frac{\partial\mathbf{p}_i}{\partial
E_{\mathbf{p}}})\phi(\mathbf{p})=0,
\end{eqnarray}
with a special solution
\begin{eqnarray}
\phi(\mathbf{p})=\frac{1}{2}\sqrt{\frac{1}{E_{\mathbf{p}}\sqrt{E_{\mathbf{p}}^2-m^2}}},
\end{eqnarray}
which is completely different from the "field" in eq.(142).

Though we have defined a time operator and find out its eigenstate,
its physical meaning is still obscure, thus let's focus on the
$\mathbf{X}$ and its eigenstate $|\mathbf{x}>$, and consider the
transition amplitude for free particles as in eq.(25)
\begin{eqnarray}
<\mathbf{x}_2,t_2|\mathbf{x}_1,t_1>=<\mathbf{x}_2|e^{-iH(t_2-t_1)}|\mathbf{x}_1>
=<0|\psiup_{\phi_1}(\mathbf{x}_2)e^{-iH(t_2-t_1)}\psiup_{\phi_1}^\dag(\mathbf{x}_1)|0>,
\end{eqnarray}
by using eqs.(142),(143) and (145), we have further
\begin{eqnarray}
<\mathbf{x}_2,t_2|\mathbf{x}_1,t_1>=<\mathbf{x}_2|e^{-i\hat{H}(t_2-t_1)}|\mathbf{x}_1>
=\int\frac{d^3\mathbf{p}}{(2\pi)^3}e^{i[\mathbf{p}(\mathbf{x}_2-\mathbf{x}_1)-E_{\mathbf{p}}(t_2-t_1)]},
\end{eqnarray}
where we have used a re-defined momentum eigenstate
\begin{eqnarray}
\overline{|\mathbf{p}>}=a^\dag(\mathbf{p})|0>=\frac{1}{\sqrt{2E_{\mathbf{p}}}}|\mathbf{p}>,\qquad
(\hat{\mathbf{p}}\overline{|\mathbf{p}>}=\mathbf{p}\overline{|\mathbf{p}>}),
\end{eqnarray}
and the following single particle completeness relation
\begin{eqnarray}
I=\int\frac{d^3\mathbf{p}}{(2\pi)^3}\overline{|\mathbf{p}><\mathbf{p}|}
=\int\frac{d^3\mathbf{p}}{(2\pi)^3}\frac{1}{2E_{\mathbf{p}}}|\mathbf{p}><\mathbf{p}|.
\end{eqnarray}
The integration in eq.(168) is a little complicated compared to
eq.(35), thus we make some approximations by expanding the phase
term
$\chi=i[\mathbf{p}(\mathbf{x}_2-\mathbf{x}_1)-E_{\mathbf{p}}(t_2-t_1)]$
near its extreme point $\partial_{\mathbf{p}}\chi=0$ which gives
\begin{eqnarray}
\mathbf{x}_2-\mathbf{x}_1=\frac{\partial
E_{\mathbf{p}}}{\partial\mathbf{p}}(t_2-t_1)=\frac{\mathbf{p}}{E_{\mathbf{p}}}(t_2-t_1)=\mathbf{v}(t_2-t_1)
=\mathbf{v}\Delta t.
\end{eqnarray}
Then we have
\begin{eqnarray}
<\mathbf{x}_2,t_2|\mathbf{x}_1,t_1>\propto\exp(-i
m\sqrt{1-\mathbf{v}^2}\Delta t)\rightarrow\exp(-i
m\int_{t_1}^{t_2}d\tau) ,
\end{eqnarray}
which is similar to eq.(36).

Here, let's make some discussions about the two fields
$\phi_1(\mathbf{x})$ and $\psiup_{\phi_1}(\mathbf{x})$. Obviously,
$\phi_1(\mathbf{x})$ is a real filed which is a scalar
representation of the Lorentz group, while
$\psiup_{\phi_1}(\mathbf{x})$ is ill defined. However, as studies
above, $\psiup_{\phi_1}(\mathbf{x})$ is useful for describing the
physics of particles, for example the eqs.(145)-(148). In other
words, it's much more like the non-relativistic Schr\"odinger filed
as analyzed in section II., so it can also be used to construct the
QM for particles by noting that $\psiup_{\phi_1}(\mathbf{x})$ can be
inserted into eq.(48) directly, thus the ensemble interpretation to
QM is still proper in the relativistic case. However, when treating
some physical processes with interactions added in , we should use
the real field $\phi_1(\mathbf{x})$, that is the QFT combining with
the anti-particle field $\phi_2(\mathbf{x})$ for which the above
analyses are still applicable.

Now, let's consider some other kinds of fields, for example the
Dirac spinor field $\psi_a(\mathbf{x})$ and the electromagnetic
field $A_\mu(\mathbf{x})$. And as we will show, for both of these
two fields, there are some problems with the position operator.
First, let's see the free Dirac field with action
\begin{eqnarray}
S=\int d^4x \bar{\psi}(i\gamma^\mu\partial_\mu-m)\psi,
\end{eqnarray}
with the field expansion[2]
\begin{eqnarray}
\psi(\mathbf{x})=\psi_1(\mathbf{x})+\psi^\dag_2(\mathbf{x})=\int
\frac{d^3\mathbf{p}}{(2\pi)^3}\frac{1}{\sqrt{2E_{\mathbf{p}}}}
\sum_s(a(\mathbf{p},s)u(p,s)e^{i\mathbf{px}}+b^\dag(\mathbf{p},s)v(p,s)e^{-i\mathbf{px}})
\\ \psi^\dag(\mathbf{x})=\psi^\dag_1(\mathbf{x})+\psi_2(\mathbf{x})=\int
\frac{d^3\mathbf{p}}{(2\pi)^3}\frac{1}{\sqrt{2E_{\mathbf{p}}}}
\sum_s(a^\dag(\mathbf{p},s)u^\dag(p,s)e^{-i\mathbf{px}}+b(\mathbf{p},s)v^\dag(p,s)e^{i\mathbf{px}}).
\end{eqnarray}
As in eq.(125), we can define another field and its conjugate
\begin{eqnarray}
\psi'(\mathbf{x})=i(\psi_1(\mathbf{x})-\psi^\dag_2(\mathbf{x}))
\\
\psi'^\dag(\mathbf{x})=-i(\psi^\dag_1(\mathbf{x})-\psi_2(\mathbf{x})),
\end{eqnarray}
and we have similarly
\begin{eqnarray}
S'=\frac{1}{2}(S_{\psi}+S_{\psi'})=\int d^4x
[\bar{\psi_1}(i\gamma^\mu\partial_\mu-m)\psi_1+\psi_2\gamma^0(i\gamma^\mu\partial_\mu-m)\psi^\dag_2],
\end{eqnarray}
that is, we separate the electron field from the positron field, and
up to the orders of the field operators, the two fields should have
the same structure, just like the case of the previous charged
scalar fields. Thus, it's also possible to define the particle
number and position operators
\begin{eqnarray}
N_{\psi_1}=\int
d^3\mathbf{x}\psi^\dag_1(\mathbf{x})\psi_1(\mathbf{x})=\int
\frac{d^3\mathbf{p}}{(2\pi)^3}\sum_sa^\dag(\mathbf{p},s)a(\mathbf{p},s)\\
\mathbf{X}_{\psi_1}=\int
d^3\mathbf{x}\psi^\dag_1(\mathbf{x})\mathbf{x}\psi_1(\mathbf{x}),
\end{eqnarray}
where we have used the normalization[2]
\begin{eqnarray}
u^\dag(p,r)u(p,s)=2E_{\mathbf{p}}\delta^{rs},\qquad
u(p,s)={\sqrt{p\cdot\sigma}\xi^s \choose
\sqrt{p\cdot\bar{\sigma}}\xi^s},(\xi^{\dag r}\xi^s=\delta^{rs}).
\end{eqnarray}
As for the position operator, after some computations we will have
\begin{eqnarray}
\mathbf{X}_{\psi_1}=\frac{i}{2}\int
\frac{d^3\mathbf{p}}{(2\pi)^3}\sum_s[a^\dag(\mathbf{p},s)\partial_{\mathbf{p}}a(\mathbf{p},s)-\partial_{\mathbf{p}}a^\dag(\mathbf{p},s)a(\mathbf{p},s)]
\nonumber\\+\frac{i}{2}\int
\frac{d^3\mathbf{p}}{(2\pi)^3}\sum_{sr}\frac{1}{2E_{\mathbf{p}}}a^\dag(\mathbf{p},s)a(\mathbf{p},r)[u^\dag(p,s)\partial_{\mathbf{p}}u(p,r)-\partial_{\mathbf{p}}u^\dag(p,s)u(p,r)],
\end{eqnarray}
where the first term is the familiar operator, while the second term
can be simplified in the following way. We rewrite the spinor in
eq.(181) in a new form which is easily to compute[10]
\begin{eqnarray}
u(p,s)=\frac{p_\mu\gamma^\mu+m}{\sqrt{E_{\mathbf{p}}+m}}u(0,s),\qquad
u(0,s)=\frac{1}{\sqrt{2}}{\xi^s \choose \xi^s},
\end{eqnarray}
then after tedious computations, the second term in eq.(182) will
become
\begin{eqnarray}
\frac{i}{2}\int
\frac{d^3\mathbf{p}}{(2\pi)^3}\sum_{sr}\frac{1}{2E_{\mathbf{p}}}a^\dag(\mathbf{p},s)a(\mathbf{p},r)K_i^{sr},\qquad
K_i^{sr}=2\xi^{\dag
s}\frac{i\epsilon_{ijk}p_j\sigma_k}{\sqrt{E_{\mathbf{p}}+m}}\xi^r
\end{eqnarray}
with the index $i$ in $K_i^{sr}$ denoted the
$\partial_{\mathbf{p}_i}$ term. Though eq.(184) is not vanishing, it
is commuting with the energy and momentum operators, then the
velocity operator and the uncertainty relation in eq.(13) are still
well defined. However, since
$\{\psi_1(\mathbf{x})_a,\psi^\dag_1(\mathbf{y})_b\}\neq
\delta_{ab}\delta^3(\mathbf{x}-\mathbf{y})$, thus the position state
can not be constructed from this field, either. Therefore, we have
to define a new "field" as\footnote{Notice that eq.(86) is in the
form of eq.(185), and we could replace the $\xi^s$ with a general
$u(0,s)$ in eq.(183).}
\begin{eqnarray}
\psiup_{\psi_1}(\mathbf{x})=\int \frac{d^3\mathbf{p}}{(2\pi)^3}
\sum_sa(\mathbf{p},s)\xi^se^{i\mathbf{px}},
\end{eqnarray}
with the communicative relation
\begin{eqnarray}
\{\psiup_{\psi_1}(\mathbf{x})_a,\psiup^\dag_{\psi_1}(\mathbf{y})_b\}=\delta_{ab}\delta^3(\mathbf{x}-\mathbf{y}),\qquad
(\sum_s\xi^s\xi^{\dag s}=I).
\end{eqnarray}
Then, we could define the position operator and its eigenstate as
\begin{eqnarray}
\mathbf{X}_{\psi_1}=\int
d^3\mathbf{x}\psiup^\dag_{\psi_1}(\mathbf{x})\mathbf{x}\psiup_{\psi_1}(\mathbf{x}),\qquad
|\mathbf{x}>_a=\psiup^\dag_{\psi_1}(\mathbf{x})_a|0>,
\end{eqnarray}
and the previous discussions for the scalar field apply here, too.
And with this ill-defined "field", we could obtain operators as
those in eqs.(145)-(149).

Now, let's consider the free electromagnetic field with action in
the vector form[10]
\begin{eqnarray}
S=\frac{1}{2}\int d^4x (\mathbf{E}^2-\mathbf{B}^2),
\end{eqnarray}
and with the Coulomb gauge $\mathbf{\nabla}\cdot\mathbf{A}=0$, we
could work completely with the following transverse field
expansions[10]
\begin{eqnarray}
\mathbf{A}(\mathbf{x})=\int
\frac{d^3\mathbf{p}}{(2\pi)^3}\frac{1}{\sqrt{2E_{\mathbf{p}}}}
\sum_{s=1}^{2}\mathbf{\epsilon}(\mathbf{p},s)(a(\mathbf{p},s)e^{i\mathbf{px}}+a^\dag(\mathbf{p},s)e^{-i\mathbf{px}})
\\ \mathbf{E}(\mathbf{x})=-\dot{\mathbf{A}}(\mathbf{x}),
\end{eqnarray}
with only two transverse components. Though the photon can be
considered as either particle or anti-particle, we could still
separate the field and the action formally, and the particle number
operator is defined as
\begin{eqnarray}
N_{\mathbf{A}}=-\frac{i}{2}\int
d^3\mathbf{x}[\dot{\mathbf{A}^\dag}(\mathbf{x})\cdot\mathbf{A}(\mathbf{x})-\dot{\mathbf{A}}(\mathbf{x})\cdot\mathbf{A}^\dag(\mathbf{x})]
\nonumber\\=\frac{1}{2}\int\frac{d^3\mathbf{p}}{(2\pi)^3}
\sum_{s=1}^{2}(a(\mathbf{p},s)a^\dag(\mathbf{p},s)+a^\dag(\mathbf{p},s)a(\mathbf{p},s)),
\end{eqnarray}
where the factor $1/2$ is due to the fact that particle and
anti-particle are the same. Thus the position operator is
\begin{eqnarray}
\mathbf{X}_{\mathbf{A}}=-\frac{i}{2}\int
d^3\mathbf{x}\mathbf{x}[\dot{\mathbf{A}^\dag}(\mathbf{x})\cdot\mathbf{A}(\mathbf{x})-\dot{\mathbf{A}}(\mathbf{x})\cdot\mathbf{A}^\dag(\mathbf{x})]
\nonumber\\=\frac{i}{2}\int\frac{d^3\mathbf{p}}{(2\pi)^3}
\sum_{s=1}^{2}(a^\dag(\mathbf{p},s)\partial_{\mathbf{p}}a(\mathbf{p},s)-\partial_{\mathbf{p}}a^\dag(\mathbf{p},s)a(\mathbf{p},s))
\nonumber\\+\frac{i}{2}\int\frac{d^3\mathbf{p}}{(2\pi)^3}\sum_{sr}a^\dag(\mathbf{p},s)a(\mathbf{p},r)
\sum_i[\mathbf{\epsilon}_i(\mathbf{p},s)\partial_{\mathbf{p}}\mathbf{\epsilon}_i(\mathbf{p},r)-\partial_{\mathbf{p}}\mathbf{\epsilon}_i(\mathbf{p},s)\mathbf{\epsilon}_i(\mathbf{p},r)],
\end{eqnarray}
which is similar to the case of Dirac field in eq.(182), and the
last term is not vanishing because of the momentum dependence of the
polarization vectors, due to the Coulomb gauge in the form
$\mathbf{p}\cdot\mathbf{\epsilon}(\mathbf{p},s)=0$. However, we
could also define a new "field" with some fixed frame in which
$\mathbf{n}\cdot\mathbf{\epsilon}(\mathbf{n},s)=0$, and an arbitrary
chosen vector $\mathbf{n}=(n_1,n_2,n_3)$
\begin{eqnarray}
\mathbf{\Lambda}(\mathbf{x})=\int \frac{d^3\mathbf{p}}{(2\pi)^3}
\sum_{s=1}^{2}\mathbf{\epsilon}(\mathbf{n},s)a(\mathbf{p},s)e^{i\mathbf{px}},
\end{eqnarray}
with the communicative relation
\begin{eqnarray}
\{\mathbf{\Lambda}(\mathbf{x})_i,\mathbf{\Lambda}^\dag(\mathbf{x})_j\}=(\delta_{ij}-n_in_j)\delta^3(\mathbf{x}-\mathbf{y}),
(\sum_s\mathbf{\epsilon}_i(\mathbf{n},s)\mathbf{\epsilon}_j(\mathbf{n},s)=\delta_{ij}-n_in_j).
\end{eqnarray}
then the particle number and position operators are
\begin{eqnarray}
N_{\mathbf{A}}=\int
d^3\mathbf{x}\mathbf{\Lambda}^\dag(\mathbf{x})\cdot\mathbf{\Lambda}(\mathbf{x})=\int\frac{d^3\mathbf{p}}{(2\pi)^3}
\sum_{s=1}^{2}a^\dag(\mathbf{p},s)a(\mathbf{p},s)\\
\mathbf{X}_{\mathbf{A}}=\int
d^3\mathbf{x}\mathbf{x}\mathbf{\Lambda}^\dag(\mathbf{x})\cdot\mathbf{\Lambda}(\mathbf{x})=\frac{i}{2}\int\frac{d^3\mathbf{p}}{(2\pi)^3}
\sum_{s=1}^{2}(a^\dag(\mathbf{p},s)\partial_{\mathbf{p}}a(\mathbf{p},s)-\partial_{\mathbf{p}}a^\dag(\mathbf{p},s)a(\mathbf{p},s)),
\end{eqnarray}
and the eigenstate of position operator is
\begin{eqnarray}
|\mathbf{x}>_i=\mathbf{\Lambda}_i^\dag(\mathbf{x})|0>,
\end{eqnarray}
although its meaning is also not clear.

In the end of this section, we try to give a somewhat systematical
study about the separation of the field, such as eq.(125). For
simplicity, we will still take the charged scalar field case. Notice
that in the action eq.(126), up to some differentials, there is a
general form of the fields
\begin{eqnarray}
\phi^\dag\phi=\phi_1^\dag\phi_1+\phi_2\phi_2^\dag+\phi_1^\dag\phi_2^\dag+\phi_2\phi_1
=(\phi_1^\dag,\phi_2)\left( \begin{array}{ccc}
1 & 1 &  \\
1 & 1 &  \\
\end{array} \right){\phi_1
\choose \phi_2^\dag}=\Phi^\dag(I+\sigma_1)\Phi,
\end{eqnarray}
similarly for the field in eq.(125)
\begin{eqnarray}
\phi'^\dag\phi'=\phi_1^\dag\phi_1+\phi_2\phi_2^\dag-\phi_1^\dag\phi_2^\dag-\phi_2\phi_1
=(\phi_1^\dag,\phi_2)\left( \begin{array}{ccc}
1 & -1 &  \\
-1 & 1 &  \\
\end{array} \right){\phi_1
\choose \phi_2^\dag}=\Phi^\dag(I-\sigma_1)\Phi,
\end{eqnarray}
that is $\phi$ and  $\phi'$ seem to be in two different "chirality"
representations! To understand these, let's start from a general
case with $\sigma_1$ replaced by $\sigma_n$. First, notice that
\begin{eqnarray}
\Phi^\dag\Phi=\phi_1^\dag\phi_1+\phi_2\phi_2^\dag ,
\end{eqnarray}
is just of the form of action in eq.(129), with
\begin{eqnarray}
\phi=\sqrt{2}u^\dag_{\uparrow_{x}}\Phi=(1,1){\phi_1 \choose
\phi_2^\dag},\qquad
(I+\sigma_1=2u_{\uparrow_{x}}u^\dag_{\uparrow_{x}}),
\end{eqnarray}
then we could define a general field
\begin{eqnarray}
\phi_{\uparrow_{n}}=\sqrt{2}u^\dag_{\uparrow_{n}}\Phi=\sqrt{2}(e^{i\frac{\varphi}{2}}\cos\frac{\theta}{2}\phi_1+e^{-i\frac{\varphi}{2}}\sin\frac{\theta}{2}\phi^\dag_2),
(I+\sigma_n=2u_{\uparrow_{n}}u^\dag_{\uparrow_{n}}),
\end{eqnarray}
and the communicative relation gives
\begin{eqnarray}
[\phi_{\uparrow_{n}}(\mathbf{x}),\phi^\dag_{\uparrow_{n}}(\mathbf{y})]
=2\{\cos^2\frac{\theta}{2}[\phi_1(\mathbf{x}),\phi_1^\dag(\mathbf{y})]-\sin^2\frac{\theta}{2}[\phi_2(\mathbf{x}),\phi_2^\dag(\mathbf{y})]\},
\end{eqnarray}
to restrict it to be zero, we have the condition $\theta=\pi/2$. In
this way, we obtain two projectors
\begin{eqnarray}
P_{\uparrow}=\frac{I+\sigma_{\theta=\frac{\pi}{2}}}{2},\qquad
P_{\downarrow}=\frac{I-\sigma_{\theta=\frac{\pi}{2}}}{2},
\end{eqnarray}
from which we have two different "chirality" representations
\begin{eqnarray}
\Phi_{\uparrow}=\frac{I+\sigma_{\theta=\frac{\pi}{2}}}{2}\Phi,\qquad
\Phi_{\downarrow}=\frac{I-\sigma_{\theta=\frac{\pi}{2}}}{2}\Phi,
\end{eqnarray}
and the action constitution will be
\begin{eqnarray}
\Phi_{\uparrow}^\dag\Phi_{\uparrow}=\Phi^\dag\frac{I+\sigma_{\theta=\frac{\pi}{2}}}{2}\Phi=\Phi^\dag
u_{\uparrow_{\theta=\frac{\pi}{2}}}u^\dag_{\uparrow_{\theta=\frac{\pi}{2}}}\Phi
=\frac{1}{2}\phi_{\uparrow_{\theta=\frac{\pi}{2}}}^\dag\phi_{\uparrow_{\theta=\frac{\pi}{2}}},
\end{eqnarray}
by using the eq.(202), similarly for other component. Easily to see,
if we further restrict $\varphi=0$, we then obtain the field $\phi$
and $\phi'$, with $i$ added in $\phi'$ to insure the hermitian for
the special real scalar field. There is a residue "chiral" symmetry
in  eq.(206) under the following transformation\footnote{The full
form in eq.(200) is symmetric under a general two-component
transformation $e^{-i\alpha\sigma_n}\Phi$, especially,
$e^{-i\alpha\sigma_3}\Phi$ is the transformation in eq.(134).}
\begin{eqnarray}
\Phi\rightarrow e^{-i\alpha\sigma_{\theta=\frac{\pi}{2}}}\Phi,
\end{eqnarray}
which induces
\begin{eqnarray}
\phi_{\uparrow_{\theta=\frac{\pi}{2}}}\rightarrow
\sqrt{2}u^\dag_{\uparrow_{\theta=\frac{\pi}{2}}}e^{-i\alpha\sigma_{\theta=\frac{\pi}{2}}}\Phi=e^{-i\alpha}\phi_{\uparrow_{\theta=\frac{\pi}{2}}},
\end{eqnarray}
the global gauge transformation! And the other component has an
opposite transformation
\begin{eqnarray}
\phi_{\downarrow_{\theta=\frac{\pi}{2}}}\rightarrow
\sqrt{2}u^\dag_{\downarrow_{\theta=\frac{\pi}{2}}}e^{-i\alpha\sigma_{\theta=\frac{\pi}{2}}}\Phi=e^{i\alpha}\phi_{\downarrow_{\theta=\frac{\pi}{2}}}.
\end{eqnarray}
Though whether this gauge transformation is the usual one is still
unknown, the above studies indeed give a systematical view on the
separations of the fields and actions.

\section{Two additive topics}

In this section, we will consider two extra topics, one is about an
operable experiment to distinguish the Copenhagen interpretation
from the ensemble one by very different experimental results, while
the other is concerned with a special ensemble state, the coherent
state.

\subsection{An Operable Experimental Test}

There are already many interpretations to the QM, for example, the
standard Copenhagen interpretation(CI), the ensemble
interpretation(EI) revived in this paper. However, it seems that all
of they are somewhat metaphysical, and one could choose any
interpretation at will. As is well known, with the CI, there is the
so called quantum collapse in the quantum measurement theory.
However, according to the EI developed previously, nothing unusual
happens. Here, we try to give an operable(quantum measurement)
experiment to test which interpretation is much more proper, via the
possible different experimental phenomena owing to the two different
interpretations.

\begin{figure}[htbp]
\setlength{\unitlength}{2mm} \centering
\includegraphics[width=3.0in]{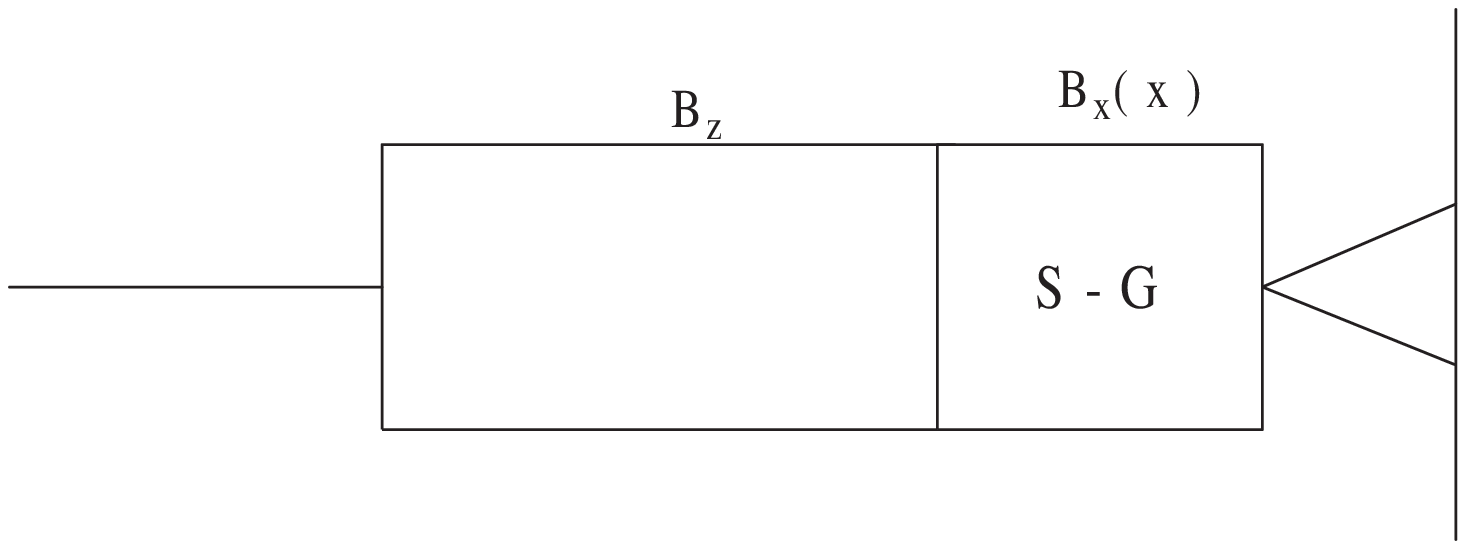}
\caption[]{The experiment sketch.}\label{fig2}
\end{figure}

We use the famous Stern-Gerlach experiment[11] as our basis, which
involves the quantum measurement of spin. Besides, we add another
apparatus to the original one, which may be considered to be a
variant of the Stern-Gerlach apparatus with the non-uniform magnetic
field replaced by a uniform one, as shown in section III.C. The
combined apparatuses are sketched in Fig.2. Now, let a beam of
electrons with specific spin $|\uparrow_x>$ travel into the first
magnetic field, obviously there's no deflections because of the
uniform magnetic field. According to QM, the final state will be
\begin{eqnarray}
|\phi>=\frac{1}{\sqrt{2}}(e^{-i\omega T}|\uparrow_z>+e^{i\omega
T}|\downarrow_z>),
\end{eqnarray}
with $\omega\equiv |e|B/2m_{e}$, and $T$ the period of the electrons
traveling in the first magnetic field. According to CI, the state in
eq.(210) is a superposition state for a single electron. Then, let
these electrons travel into the second magnetic field, i.e. making
measurements on the spins, and the Stern-Gerlach apparatus would
split the beam into two distinct components. According to CI, the
probability for observing the $|\uparrow_x>$ is
\begin{eqnarray}
|<\uparrow_x|\phi>|^2=\cos^2\omega T,
\end{eqnarray}
similarly for $|\downarrow_x>$,
\begin{eqnarray}
|<\downarrow_x|\phi>|^2=\sin^2\omega T.
\end{eqnarray}

However, according to EI, we should treat the state in eq.(210) as
an ensemble state, which briefly says that the electrons in the beam
are roughly divided into two parts with almost the same particle
number, one part with state $|\uparrow_z>$, while the other
$|\downarrow_z>$, and with $e^{\pm i\omega T}$ some irrelevant phase
terms\footnote{The phase terms could have their effects in some
physical process, for example the Young's double-slit experiment in
the section III.E. Here, the measurement is about spin, so those
phase terms may have no effects. }. This means that, when we make
measurements with the Stern-Gerlach apparatus, the probability for
observing the $|\uparrow_x>$ will be
\begin{eqnarray}
|<\uparrow_z|\phi>|^2
|<\uparrow_x|\uparrow_z>|^2+|<\downarrow_z|\phi>|^2
|<\uparrow_x|\downarrow_z>|^2=\frac{1}{2},
\end{eqnarray}
with the same probability for $|\downarrow_x>$.

Since in the experiment, the probabilities are relevant to the
ratios of particle numbers, then there are obvious differences
between CI and EI, comparing eqs.(211)(212) with (213). That is, by
tuning the strength of magnetic filed or the length of the first
apparatus, we could vary the phases in eqs.(211) and (212) so as to
obtain alterable probabilities, correspondingly the particle numbers
in the two components. We even could obtain nothing in one of the
two components when
\begin{eqnarray}
\omega T=\frac{1}{2}n\pi \qquad n=0,\pm 1,\pm 2,\ldots.
\end{eqnarray}
However, according to EI, there are always two components with equal
particle numbers up to some admissible experimental errors.
Therefore, with the possible results of the above experiment, we
could obtain the following conclusions:

(1)\emph{The notable differences between the particle numbers are
observed by tuning the possible parameters, then CI is more proper,
and furthermore, we obtain the strict evidences of the superposition
state.}

(2)\emph{The particle numbers of the two components are always equal
up to admissible experimental errors, this means that CI is wrong,
while EI is more proper, and the state in eq.(210) is not a
superposition state.}

(3)\emph{We observe a complete new phenomenon which can not be
explained by either CI or EI, then we have to find out another
proper interpretation to QM.}

Although the CI to QM is familiar to us, there are still some
corrections, which can be seen by adding the wave functions to the
state in eq.(210)\footnote{Notice that the state in eq.(215) is of
the result of eq.(92) with the operator in eq.(217) acting on the
state. }
\begin{eqnarray}
\frac{1}{\sqrt{2}}(e^{i[pL-(E+\omega)T]}|p,\uparrow_z>+e^{i[p'L-(E'-\omega)T']}|p',\downarrow_z>),\qquad
E+\omega=\bar{E}=E'-\omega,
\end{eqnarray}
with $L$ the length of the first apparatus and $\bar{E}$ the energy
of the electron before entering into the first apparatus. In short,
the above function can be rewritten as
\begin{eqnarray}
\frac{1}{\sqrt{2}}(|E+\omega,p,\uparrow_z>+e^{i\delta}|E'-\omega,p',\downarrow_z>),
\end{eqnarray}
with the meaning of superposition of two different states. To obtain
the phase terms in eq.(215), we have used a space-time translation
\begin{eqnarray}
\exp i\Delta t(\dot{X}P-H),
\end{eqnarray}
with the velocity operator defined in eq.(19), for the electrons are
still free within a constant potentials $V\sim \pm \omega$.

From eq.(215), we can see that the complete probabilities
corresponding to eqs.(211) and (212) will be more complicated due to
$p'$ and $T'$. However, if the frequency $\omega$ is much smaller
than the energy of the electron, i.e. $\omega\ll \bar{E}$, we can
obtain the corrections of eqs.(211) and (212) up to first order.
These can be seen as follows by computing the phase difference
classically
\begin{eqnarray}
(p-p')L-\bar{E}(T-T')\approx-3 \omega \bar{T},
\end{eqnarray}
with $\bar{T}\simeq Lm_e/\bar{p}$, $\bar{p}\simeq
\sqrt{2m_e\bar{E}}$. Then the probabilities in eqs.(211) and (212)
will approximately be
\begin{eqnarray}
\cos^2\frac{3}{2}\omega \bar{T}\qquad \sin^2\frac{3}{2}\omega
\bar{T}.
\end{eqnarray}
We can even construct the real wave-packet for each state
\begin{eqnarray}
\int dk\phi(k)e^{i[kL-(E+\omega)T]}\qquad \int
dk'\phi(k')e^{i[k'L-(E'-\omega)T']},
\end{eqnarray}
with the respective probability densities $|\phi(k)|^2$ and
$|\phi(k')|^2$ which are mainly valued near $k=p$ and $k'=p'$.
Furthermore, notiing that the average momentum $\bar{k}$ should be
constant in the process, then we have
\begin{eqnarray}
\phi(k)=\phi(k')e^{i\delta},
\end{eqnarray}
and the phase term can simply be ignored. With these, eq.(218) is
still valid. In fact, within the CI, since the phase terms always
affect the probabilities, the results of the experiment are always
tenable. As for EI, the phase terms are irrelevant to the physical
results, so the conclusions above are unchanged.

\subsection{Coherent State: From Quantum Field to Classical Field}

In the quantization of oscillator, there is a special state, the so
called coherent state defined as
\begin{eqnarray}
|z>=e^{-\frac{1}{2}|z|^2}\sum_{n=0}^{\infty}\frac{z^n}{\sqrt{n!}}|n>,
\qquad <z|z>=1.
\end{eqnarray}
Treating the oscillator as an 1-dimensional field, then coherent
state is a superposition of infinitely many states with different
particle numbers, and according to section III., it's actually an
ensemble state. Besides, this state can be related to the classical
oscillator via
\begin{eqnarray}
a|z>=z|z>.
\end{eqnarray}
In this subsection, we extend this coherent state to the general
field case, including the Bosonic and Fermionic cases, indicating
that the classical field could be obtained from the the quantized
field with the filed operator acting on the extended coherent
states.

For Bosonic case, we take the scalar field as examples, and since
the communicative relation
$[a(\mathbf{p}),a^\dag(\mathbf{p})]=(2\pi)^3\delta^3(\mathbf{0})$ is
a divergence, then re-normalizing these operators by a factor
$1/\sqrt{(2\pi)^3\delta^3(\mathbf{0})}$ and discreting them
formally, so that we have
\begin{eqnarray}
[a_{\mathbf{p}},a^\dag_{\mathbf{q}}]=\delta_{\mathbf{p}\mathbf{q}}.
\end{eqnarray}
Now, the conditions are almost the same as those of the oscillator
case, then we could define the coherent state for each momentum
state $\mathbf{p}$ as
\begin{eqnarray}
|\phi(\mathbf{p})>=e^{-\frac{1}{2}|\phi(\mathbf{p})|^2}\sum_{n=0}^{\infty}\frac{\phi(\mathbf{p})^n}{\sqrt{n!}}(a^\dag_{\mathbf{p}})^n|0>,
\qquad <\phi(\mathbf{p})|\phi(\mathbf{p})>=1.
\end{eqnarray}
For different momentum states, we have
\begin{eqnarray}
<\phi(\mathbf{q})|\phi(\mathbf{p})>=e^{-\frac{1}{2}|\phi(\mathbf{p})|^2-\frac{1}{2}|\phi(\mathbf{q})|^2},(\mathbf{p}\neq
\mathbf{q}).
\end{eqnarray}
Further, like eq.(223), we have
\begin{eqnarray}
\left\{ \begin{array}{ll}
a_{\mathbf{p}}|\phi(\mathbf{p})>=\phi(\mathbf{p})|\phi(\mathbf{p})> &\\
a_{\mathbf{p}}|\phi(\mathbf{q})>=0 & {(\mathbf{p}\neq\mathbf{q})}
\end{array} \right..
\end{eqnarray}
Now we collect all the momentum states altogether, and define a
state
\begin{eqnarray}
|\psiup>=\prod_{\mathbf{p}}|\phi(\mathbf{p})>,
\end{eqnarray}
with normalization
\begin{eqnarray}
<\psiup|\psiup>=\prod_{\mathbf{p},\mathbf{q}}<\phi(\mathbf{q})|\phi(\mathbf{p})>=\prod_{\mathbf{p}}<\phi(\mathbf{p})|\phi(\mathbf{p})>=1,
\end{eqnarray}
and the equation
\begin{eqnarray}
a_{\mathbf{p}}|\psiup>=\prod_{\mathbf{q}}a_{\mathbf{p}}|\phi(\mathbf{q})>=\prod_{\mathbf{q}}\phi(\mathbf{p})|\phi(\mathbf{q})>=\phi(\mathbf{p})|\psiup>,
\end{eqnarray}
then the classical filed derived from the quantized one is
\begin{eqnarray}
<\psiup|\hat{\psi}(\mathbf{x})|\psiup>=<\psiup|\sum_{\mathbf{p}}a_{\mathbf{p}}e^{i\mathbf{p}\mathbf{x}}|\psiup>=\sum_{\mathbf{p}}\phi(\mathbf{p})e^{i\mathbf{p}\mathbf{x}}=\psi(\mathbf{x}).
\end{eqnarray}

Now, let's consider the Fermionic case, the first step is the same
as that of the Bosonic case, that is the discretion of the operators
\begin{eqnarray}
\{a_{\mathbf{p}},a^\dag_{\mathbf{q}}\}=\delta_{\mathbf{p}\mathbf{q}}.
\end{eqnarray}
Then we have to define the coherent state, unlike the Bosonic case
where $\phi(\mathbf{p})$ is a c-number, here, we should deal with
Grassmann numbers[2] satisfying $\phi_1\phi_2=-\phi_2\phi_1$,
further the complex conjugate is defined as
\begin{eqnarray}
(\phi_1\phi_2)^*\equiv\phi^*_2\phi^*_1 =-\phi^*_1\phi^*_2.
\end{eqnarray}
Thus we can define the coherent state as
\begin{eqnarray}
|\phi(\mathbf{p})>=e^{-\frac{1}{2}\phi^*\phi(\mathbf{p})}(1+\phi(\mathbf{p})a^\dag_{\mathbf{p}})|0>,
\end{eqnarray}
with normalization
\begin{eqnarray}
<\phi(\mathbf{p})|\phi(\mathbf{p})>=e^{-\phi^*\phi(\mathbf{p})}(1+\phi^*\phi(\mathbf{p}))=1,
\end{eqnarray}
where we have used the relations
$e^{-\phi^*\phi(\mathbf{p})}=1-\phi^*\phi(\mathbf{p})$ and
$\phi^*\phi\phi^*\phi=-\phi^*\phi^*\phi\phi=0$. Then with the
annihilator acting on the state, we have
\begin{eqnarray}
a_{\mathbf{p}}|\phi(\mathbf{p})>=e^{-\frac{1}{2}\phi^*\phi(\mathbf{p})}\phi(\mathbf{p})|0>=\phi(\mathbf{p})|0>,
\end{eqnarray}
or
\begin{eqnarray}
<\phi(\mathbf{p})|a_{\mathbf{p}}|\phi(\mathbf{p})>=\phi(\mathbf{p}).
\end{eqnarray}
As the Bosonic case, we could also define a state by noting that for
different momentum states, the $|\phi(\mathbf{p})>$'s are commuting
\begin{eqnarray}
|\psiup>=\prod_{\mathbf{p}}|\phi(\mathbf{p})>,\qquad
<\psiup|\psiup>=1,
\end{eqnarray}
and
\begin{eqnarray}
<\psiup|a_{\mathbf{p}}|\psiup>=\prod_{\mathbf{q}\neq\mathbf{p}}<\phi(\mathbf{q})|\phi(\mathbf{q})><\phi(\mathbf{p})|a_{\mathbf{p}}|\phi(\mathbf{p})>=\phi(\mathbf{p}).
\end{eqnarray}
With these, we then have the classical field, ignoring some spinor
structures
\begin{eqnarray}
<\psiup|\hat{\psi}(\mathbf{x})|\psiup>=<\psiup|\sum_{\mathbf{p}}a_{\mathbf{p}}e^{i\mathbf{p}\mathbf{x}}|\psiup>=\sum_{\mathbf{p}}\phi(\mathbf{p})e^{i\mathbf{p}\mathbf{x}}=\psi(\mathbf{x}).
\end{eqnarray}

The reason for the direct product structure of the state $|\psiup>$
for both cases is mainly because that different momentum states are
independent from each other in the free field case. If not, there
would be some states such as $a|1>+b|1,2>$, in which different
states are interrelated with each other, so that the probability in
eq.(48) is not valid. However, with the direct product structure, we
can still have
\begin{eqnarray}
P_{\mathbf{p}}= \frac{<\psiup|a_{\mathbf{p}}^\dag
a_{\mathbf{p}}|\psiup>}{<\psiup|\sum_{\mathbf{p}}a_{\mathbf{p}}^\dag
a_{\mathbf{p}}|\psiup>}=\frac{\phi^*(\mathbf{p})\phi(\mathbf{p})}{\sum_{\mathbf{p}}\phi^*(\mathbf{p})\phi(\mathbf{p})},
\end{eqnarray}
and for Bosonic case, its meaning is easy to understand, with
$|\phi(\mathbf{p})|^2$ treated as some classical intensity strength,
while for the Fermionic case unclear. In fact, the state in eq.(234)
is not a real ensemble state in the usual sense, due to the
character of Grassmann numbers. And a real ensemble state for the
Fermionic case should be of the form
\begin{eqnarray}
|\phi(\mathbf{p})>=(1+|\phi(\mathbf{p})|^2)^{-1/2}(1+\phi(\mathbf{p})a^\dag_{\mathbf{p}})|0>,
\end{eqnarray}
with the c-number $\phi(\mathbf{p})$, but if so, we would not obtain
the classical anticommuting fields.

\section{Conclusions and Dissuasions}

In this paper, we develop in details a new approach to the QM. From
the non-relativistic Schr\"odinger field theory, the main three
approaches to QM are obtained consistently, the Schr\"odinger
equation (2) as field equation, the Heisenberg equations (10) and
(11) for the momentum and position operators of the particles, and
the Feynman path integral formula eq.(25). With the identity of
eqs.(47) and (48), the probability concepts of QM can be induced
from the statistical properties of some collection of particles,
with the use of concepts of ensemble states, such as the states in
eqs.(62) and (64). Therefore, the modifications to the SQM is
inevitable, for example, the Schr\"odinger equation (74) which is
believed to be fundamental in SQM can be derived from general
quantized field equation (75). The most important modification is
the concept of superposition state which in our view belong to a
class of states with the form of eqs.(81),(83) and (84), while the
rest are almost the ensemble states. Then, the quantum collapse in
SQM measurement is just misunderstanding, and the EPR paradox is
also solved in eqs.(96) and (97). In addition, the most famous
experiment, the double-slit interference experiment is interpreted
in field theoretical languages, too, with the particle number
distribution eq.(118) obtained.

When considering the relativistic field theory, a method of
separating the particle field from the anti-particle field is
developed in eqs.(124)-(129), so that the operators which are
physical observables of particles are possible to be defined, see
for examples, eqs.(145)-(148). This method is useful for the scalar
field well, while for the Dirac field and the gauge field, there are
some problems with their position operators, and to resolve them, we
introduce some ill-defined "fields" so that the ensemble
interpretation is still proper. An operable experiment is proposed
in section V.A. to distinguish the Copenhagen interpretation from
the ensemble one via very different experimental results, see
eqs.(211)-(214). We also make some extensions of the concepts of
coherent state for the oscillator to both the Bosonic and Fermionic
fields, obtaining the corresponding classical fields in eqs.(231)
and (240).

Now, let's make some general discussions, especially about the
differences between the QM in the standard form and the one derived
from the QFT, on the framework of the derivations in sections II.
and IV..First, let's list some familiar rules about the standard QM,

\emph{1.The states of particles or systems are described by the wave
function formulism or the Dirac's bra-ket formalism in Hilbert
space. The most familiar and important states of a single particle
are its positions $|\mathbf{x}>$ and momentum $|\mathbf{p}>$.}

\emph{2. There are single particle operators which are some physical
observables whose eigenvalues can be measured in experiments, for
examples the energy $\hat{H}$, the momentum $\hat{\mathbf{p}}$, the
position $\hat{\mathbf{x}}$ for single particle, and the
communicative relations among them.}

\emph{3.The state of the system $|\phi(t)>$ satisfies the time
evolution equation (74).}

For the non-relativistic case, the above three rules are perfectly
realized, which are described well in field theoretical languages in
section II., with the field operators in eqs.(5)-(7), and the state
in eq.(20), we can further induce the single particle operators
\begin{eqnarray}
\hat {H}=\frac{\hat{\mathbf{p}}^2}{2m}+V(\hat{\mathbf{x}}),\qquad
\hat {\mathbf{p}}=-i\mathbf{\nabla},\qquad
\hat{\mathbf{x}}=\mathbf{x},
\end{eqnarray}
and further the mean value of some operator $\hat{O}$ within some
state $|\phi>$ is
\begin{eqnarray}
<\phi|\int d^3\mathbf{x}\psi^\dag(\mathbf{x})\hat
{O}\psi(\mathbf{x})|\phi>.
\end{eqnarray}
If $|\phi>=\sum_n\alpha_n|n>,|n>=a^\dag_n|0>,\sum_n|\alpha_n|^2=1$,
then we have
\begin{eqnarray}
\psi(\mathbf{x})|\phi>=\sum_na_n\psi_n(\mathbf{x})|\phi>=\sum_n\alpha_n\psi_n(\mathbf{x})|0>=\phi(\mathbf{x})|0>,
\end{eqnarray}
from which we obtain the wave function
$\phi(\mathbf{x})$\footnote{Don't confuse with the classical field
in eq.(231). }, then eq.(244) reduces to
\begin{eqnarray}
\int d^3\mathbf{x}\phi^*(\mathbf{x})\hat {O}\phi(\mathbf{x}),
\end{eqnarray}
which is the familiar QM formalism, specially for $\hat{O}=I$, the
probability assumption in QM, which is $1$ in this case, confirming
the eqs.(47)-(49).

However, for the relativistic case, the above three rules are not
always proper, even for the free field case. As shown in section
IV., we find out a method to separate the particle field from the
anti-particle filed, then one may simply believe that the above
rules should be satisfied, too. For the Dirac field, rules 2. and 3.
are realized, while for the first one, the position state is not
well defined. The single particle operators for Dirac field are
\begin{eqnarray}
\hat {H}=\gamma^0\gamma^i\hat{\mathbf{p}}+m\gamma^0,\qquad \hat
{\mathbf{p}}=-i\mathbf{\nabla},\qquad \hat{\mathbf{x}}=\mathbf{x}.
\end{eqnarray}
However, for the scalar and electromagnetic fields, there are not
single particle operators formally in the original filed formula,
because of the twice differentials about time. These can also be
seen in the following way, supporting a state $|\phiup>$
\begin{eqnarray}
|\phiup>=\int
\frac{d^3\mathbf{p}}{(2\pi)^3}\frac{\beta(\mathbf{p})}{\sqrt{2E_{\mathbf{p}}}}|\mathbf{p}>,
\qquad <\phiup|\phiup>=\int
\frac{d^3\mathbf{p}}{(2\pi)^3}|\beta(\mathbf{p})|^2=1,
\end{eqnarray}
then for the scalar particle field $\phi_1$ in eq.(124), eq.(245)
will be
\begin{eqnarray}
\phi_1(x)|\phiup>=\int
\frac{d^3\mathbf{p}}{(2\pi)^3}\frac{\beta(\mathbf{p})}{\sqrt{2E_{\mathbf{p}}}}e^{i\mathbf{p}\mathbf{x}}|0>,
\end{eqnarray}
which is not normalized to $1$, due to the factor
$1/\sqrt{2E_{\mathbf{p}}}$. This is also the case for the
electromagnetic field. There are also some other problems, for
examples the energy and momentum operators of the scalar field
defined in eqs.(131)-(133) are completely different from the
non-relativistic case formally, but they are all resulting from the
principle of space-time transformations.

Therefore, to achieve the above three rules, we have to define some
ill-defined "fields" as in eqs.(142),(185) and (193), which are
similar to the non-relativistic field. And with these "fields", we
can also introduce single particle operators with the forms in
eq.(149), then the above three rules of QM are all satisfied,
especially the wave functions are well defined, for example for the
electromagnetic field(or photon) case, we can extend the state
$|\phiup>$ by including the polarizations
\begin{eqnarray}
|\phiup>=\int
\frac{d^3\mathbf{p}}{(2\pi)^3}\sum_s\frac{\beta(\mathbf{p},s)}{\sqrt{2E_{\mathbf{p}}}}|\mathbf{p},s>,
\qquad <\phiup|\phiup>=\int
\frac{d^3\mathbf{p}}{(2\pi)^3}\sum_s|\beta(\mathbf{p},s)|^2=1,
\end{eqnarray}
then the wave function will be from
\begin{eqnarray}
\mathbf{\Lambda}_i(\mathbf{x})|\phiup>=\int
\frac{d^3\mathbf{p}}{(2\pi)^3}\sum_s\beta(\mathbf{p},s)\mathbf{\epsilon}_i(\mathbf{n},s)e^{i\mathbf{p}\mathbf{x}}|0>,
\end{eqnarray}
with normalization
\begin{eqnarray}
<\phiup|\int
d^3\mathbf{x}\mathbf{\Lambda}^\dag(\mathbf{x})\cdot\mathbf{\Lambda}(\mathbf{x})|\phiup>=\int
\frac{d^3\mathbf{p}}{(2\pi)^3}\sum_s|\beta(\mathbf{p},s)|^2=1.
\end{eqnarray}
Though the QM with standard form is realized with those ill-defined
"field", the Lorentz group is broken owing to the bad transformation
properties of those "fields", just like the non-relativistic case.
Therefore, when considering some general physical processes which
should be Lorentz invariant, the standard form of QM in which single
particle operators can be defined, is not enough and even wrong,
instead QFT is the most proper description. In this sense, the
position operators which can be defined well with the ill-defined
"field", as in eqs.(148),(187) and (196), together with the
corresponding position eigenstates are actually not real physical,
and the only physical observables are all those which can be
obtained from the invariance of the action under some
transformations, for examples, the energy and momentum, the charge
and the particle number with transformations in eq.(134).

Here is a note about the relationships between the single particle
operators, defined in eqs.(149), (243) and (247), and the
corresponding ones constructed with fields, for example the
operators in eqs.(5)-(8) for the non-relativistic case, and those in
eqs.(132),(133),(135) and (136) for the scalar field case, or those
defined in eqs.(145)-(148), with the use of ill-defined "fields".
For the non-relativistic case, the communicative relations among the
operators constructed with fields seem to be determined completely
by the structure of the single particle operators, as long as the
field communicative relations in eq.(3) for both Bosonic and
Fermionic cases are imposed, so do the operators defined with the
ill-defined "fields", with the communicative relations among these
ill-defined "field" satisfied, such as those in eqs.(143), (186) and
(194). However, for the well-defined relativistic fields, the above
structure is not always enough. This can be seen generally as
follows. Consider two single particle operators $\hat{O}_1$ and
$\hat{O}_1$, which in field theory may be of the forms
\begin{eqnarray}
O_1=\int
d^3\mathbf{x}\phi^\dag(\mathbf{x})\hat{O}_1\phi(\mathbf{x}),\qquad
O_2=\int
d^3\mathbf{x}\phi^\dag(\mathbf{x})\hat{O}_2\phi(\mathbf{x}),
\end{eqnarray}
then the problem is
\begin{eqnarray}
[O_1,O_2]\stackrel{?}{=}\int
d^3\mathbf{x}\phi^\dag(\mathbf{x})[\hat{O}_1,\hat{O}_2]\phi(\mathbf{x}),
\end{eqnarray}
which is obviously true for the non-relativistic field case and
ill-defined "field" case, but not for all the relativistic fields
generally. Taking the charged scalar field as example , if
$\phi(\mathbf{x})$ is treated as the full field, then with
$[\phi(\mathbf{x}),\phi\dag(\mathbf{y})]=0$, the left hand side of
eq.(254) is $0$ identically, while the right hand side is not. If
$\phi(\mathbf{x})$ is only as the particle field, and considering
the operators $\hat{\mathbf{p}}$ and $\hat{\mathbf{x}}$, then after
some computations, we have
\begin{eqnarray}
[\mathbf{X}_i,\mathbf{P}_j]=\int
\frac{d^3\mathbf{p}}{(2\pi)^3}(\frac{1}{2E_{\mathbf{p}}})^2(\delta_{ij}-\frac{\mathbf{p}_i\mathbf{p}_j}{E_{\mathbf{p}}^2})a^\dag(\mathbf{p})a(\mathbf{p}),
\end{eqnarray}
which is completely different from $\int
d^3\mathbf{x}\phi^\dag(\mathbf{x})[\hat{\mathbf{x}}_i,\hat{\mathbf{p}}_j]\phi(\mathbf{x})$.
In fact in this case, the momentum operator is not of the above form
at all, but should be constructed with the canonical momentum field
$\pi(\mathbf{x})$, as in eq.(133), so do other operators, since
$[\phi(\mathbf{x}),\pi(\mathbf{y})]=i\delta^3(\mathbf{x}-\mathbf{y})$
for the whole field (including both the particle and anti-particle
field). From these, we can conclude that, in general, the single
particle operators in QM are not enough to determine the structure
of the operators constructed with fields. In fact, we can construct
a lot of operators with the fields, the space-time coordinates and
their differentials, among which only a few have some physical
meanings, i.e. those which can be derived from the symmetries of the
actions under some transformations. In this sense, the single
particle operators are not fundamental, but instead, the quantized
fields are! Even, we can treat the single particle operators as just
the induced formal results of the corresponding field operators
resulting from the transformations, of course, the position operator
is not of this kind, and it doesn't exit physically at all.

However, just like the non-relativistic field case, there are indeed
some special fields and some special cases, where eq.(254) is true.
The operators in eq.(247) for the Dirac field are of this special
kind, owing to its spinor structure, and their communicative
relations can determine the whole structure via the communicative
relation of the full Dirac field
\begin{eqnarray}
\{\psi(\mathbf{x})_a,\psi^\dag(\mathbf{y})_b\}=\delta_{ab}\delta^3(\mathbf{x}-\mathbf{y}).
\end{eqnarray}
Taking the position operator $\int
d^3\mathbf{x}\psi^\dag(\mathbf{x})\mathbf{x}\psi(\mathbf{x})$, for
example, which is in fact not a real position operator for it has no
corresponding eigenstate because of $\psi(\mathbf{x})|0>\neq0$, and
with eq.(256), we could obtain the velocity operator, $\int
d^3\mathbf{x}\psi^\dag(\mathbf{x})\gamma^0\gamma^i\psi(\mathbf{x})$,
which in the sense of single particle operator, can also be derived
from the equation
\begin{eqnarray}
\dot
{\hat{\mathbf{x}}}^i=i[\hat{H},\hat{\mathbf{x}}^i]=\gamma^0\gamma^i,
\end{eqnarray}
i.e. eq.(254) is realized in this case. Even for the electron field
$\psi_1(\mathbf{x})$ in eq.(174), the eq.(257) is still proper,
since the commutating of position operator in  eq.(182) with the
corresponding energy operator is just the velocity operator, which
can also be verified directly with $\psi_1(\mathbf{x})$ substituted,
after some tedious computations.

With the above discussions about the single particle operators,
let's study generally the statistical properties of the ensemble
which were briefly exhibited in section II.A., such as eqs.(48) and
(49). For simplicity, we work still in the non-relativistic case.
For a general operator $O=\int
d^3\mathbf{x}\psi^\dag(\mathbf{x})\hat {O}\psi(\mathbf{x})$, a
single particle ensemble state $|\phi>$ contains almost all the
statistical information about some specific property(such as the
energy state) of the single particle system, for example the
expectation value $<\phi|O|\phi>$, which can also be considered to
be the mean value of a collection of particles which realize that
ensemble. Furthermore, when considering the fluctuations, we need
the expectation value of $O^2$. After some computations, we have
\begin{eqnarray}
O^2=\int d^3\mathbf{x}\psi^\dag(\mathbf{x})\hat
{O}^2\psi(\mathbf{x})\mp \int
d^3\mathbf{x}\psi^\dag(\mathbf{x})\{\int
d^3\mathbf{y}\psi^\dag(\mathbf{y})\hat {O_y}\psi(\mathbf{y})\}\hat
{O_x}\psi(\mathbf{x}),
\end{eqnarray}
where the communicative relations in eq.(3) are used. Then for the
single particle ensemble state $|\phi>$, the last term vanishes by
using eq.(245), and QM formula is fulfilled, and the fluctuations
for single particle can be derived in the familiar way. However, for
a N-particle ensemble, the last term in eq.(258) will not vanish,
since there are correlations among those particles in the N-particle
system. We can see these with a simple example, for instance, the
energy operator together with its eigenstates, obviously, for this
case, $<H^2>$ is $\overline{E_t^2}$, that is the expectation value
of the square of the total energy, while the first term in eq.(258)
is $\overline{\sum_i^{N}E_i^2}$, which lacks the correlations
between different energy states. This is easy to understand, by
noting that the N-particle system is as a whole just like a single
particle. Therefore, QFT is much useful than QM when treating the
many-particle systems, and also in this sense, QFT is a fundamental
theory.

Though there may be some special cases, the QM with the standard
form of the above three rules is indeed not a fundamental theory
generally, not only because QM can be consistently derived from QFT
both non-relativistically and relativistically, but also because of
the non-universality of those assumed rules as a general
quantization scheme, for we could not measure or determine
theoretically the physical states of the whole(or global) field in
general\footnote{In the standard quantization scheme, the field
operator, just like the position operator, should satisfy the
eigenvalue equation
$\hat{\phi}(\mathbf{x})|\phi>=\phi(\mathbf{x})|\phi>$[2], but it
seems impossible to realize physically. We can (classically) measure
exactly the static field, such as the electrostatic field, with a
test particle, by observing the motion of the particle, but not
possible for a general dynamical field.}, but only describe them
formally mathematically. What we can obtain or measure are only the
states of the particles excited from those fields, and the
corresponding local properties. Thus in this sense, QFT is the
unique fundamental theory in principle, in which fields are
fundamental elements of our physical world, in the nowadays
experimental limit.

\begin{acknowledgments}
\indent\indent This work is supported by NSF (10703001) and the
Fundamental Research Funds for the Central Universities (DUT10LK31).
The author is very grateful to his family, and his friends,
especially He Ma, for their encouragements.
\end{acknowledgments}
Email address:

\textasteriskcentered flyphys@mail.dlut.edu.cn

\end{document}